\documentstyle[12pt,aasms4]{article}
\newcommand{\eg}{e.g., }
\newcommand{\ie}{i.e., }

\newcommand{\etal}{et~al.\ }
\newcommand{\msun}{\mbox{$M_\odot$}}
\newcommand{\ha}{{\it H$\alpha$}~}

\newcommand{\mm}{{\it $\mu$m}~}
\newcommand{\xx}{$\times$}

\newcommand{\ksm}{km s$^{-1}$ Mpc$^{-1}$~}

\begin{document}

\title{A {\it Hubble Space Telescope}\footnote{Based on observations made with
the NASA/ESA {\it Hubble Space Telescope}, obtained at the Space Telescope
Science Institute, which is operated by the Association of Universities for
research in Astronomy, Inc., under NASA contract NAS 5-26555.} Imaging Survey
of Nearby Active Galactic Nuclei} 

\author{Matthew A. Malkan, Varoujan Gorjian, and Raymond Tam}
\authoraddr{University of California, Los Angeles \\
            Department of Physics \& Astronomy \\
            8971 Mathematical Sciences Bldg. \\
            Los Angeles, CA 90095-1562, USA}
\affil{Department of Physics \& Astronomy, University of California,\\
Los Angeles, CA 90095-1562\\
malkan@astro.ucla.edu, vg@astro.ucla.edu, utam@physics.ucla.edu}

\begin{abstract}
We have obtained  WFPC2 images of 256 of the nearest (z$\leq$0.035) Seyfert 1,
Seyfert 2, and starburst galaxies. Our 500-second broadband (F606W) exposures
reveal much fine-scale structure in the centers of these galaxies, including
dust lanes and patches, bars, rings, wisps and filaments, and tidal features
such as warps and tails. Most of this fine structure cannot be detected in
ground based images. We have assigned qualitative classifications for these
morphological features, a Hubble type for the inner region of each galaxy, and
also measured quantitative information such as 0.18 and 0.92 arcsecond
aperture magnitudes, position angles and ellipticities where possible. 

There is little direct evidence for unusually high rates of interaction in the
Seyfert galaxies.  Slightly less than 10\% of all the galaxies show tidal
features or multiple nuclei. The incidence of inner starburst rings is about
10\% in both classes of Seyfert galaxies. In contrast, galaxies with H II
region emission line spectra appear substantially more irregular and clumpy,
because of their much higher rates of current star formation per unit of
galactic mass. 

The presence of an unresolved central continuum source in our {\it HST} images
is a virtually perfect indicator of a Seyfert 1 nucleus as seen by
ground-based spectroscopy. Fifty-two percent (52\%) of these Seyfert 1 point
sources are saturated in our images; we use their wings to estimate magnitudes
ranging from 15.8 to 18.5. The converse is not universally true, however, as
over a third of Seyferts with direct spectroscopic evidence for broad Balmer
wings show no nuclear point source. These 34 resolved Seyfert 1's have fainter
nonstellar nuclei, which appear to be more extinguished by dust absorption.
Like the Seyfert 2's, they have central surface brightnesses consistent with
those expected for the bulges of normal galaxies. 

The rates for the occurrences of bars in Seyfert 1's and 2's and non-Seyferts
are the same. We found one significant morphological difference between the
host galaxies of Seyfert 1 and Seyfert 2 nuclei. The Seyfert 2 galaxies are
significantly more likely to show nuclear dust absorption, especially in 
lanes and patches which are irregular or reach close to the nucleus. A few
simple tests show that the difference cannot be explained by different average
redshifts or selection techniques. It is confirmed by our galaxy morphology
classifications, which show that Seyfert 1 nuclei reside in earlier type
galaxies than Seyfert 2 nuclei. If, as we believe, this is an intrinsic
difference in host galaxy properties, it would undermine one of the postulates
of the strong unification hypothesis for Seyfert galaxies, that they merely
appear different due to the orientation of their central engine. The excess
galactic dust we see in Seyfert 2's may cause substantial absorption which
obscures their hypothesized broad-emission-line regions and central nonstellar
continua. This galactic dust could produce much of the absorption in Seyfert 2
nuclei which had instead been attributed to a thick dusty accretion torus
forming the outer part of the central engine. 

\end{abstract} 


\section{Introduction}

Several causal connections have been proposed between an active galactic
nucleus (AGN) and the host galaxy in which it resides. The principal ways
in which the latter could affect the former are through influencing a) the
formation of a nonstellar central engine; b) its fueling; and c) obscuring it 
from our view, (which can alter the central engine's appearance even if it is 
not physically affected.)

It is widely believed that active galactic nuclei (AGNs) are powered by
non-spherical accretion onto massive black holes. This is partly because this
model has the lowest fuel supply requirements: an AGN's luminosity is
proportional to its mass accretion rate, which would be about 0.01 \msun
year$^{-1}$ for a bright Seyfert nucleus. It is not known how this rate of
fuel supply can be brought from the host galaxy down to several thousand
Schwarzschild radii (of order 10$^{17}$ cm for a ``typical" Seyfert galaxy
black hole mass of 10$^8$ \msun (Malkan \markcite{a60} 1983)) at which point
viscous processes are supposed to drive the final accretion onto the black
hole.  One speculation is that a close interaction with another galaxy can
distort the galactic potential and disturb the orbits of gas clouds
sufficiently to carry a significant mass of fuel into the galaxy's center
(Shlosman \etal\markcite{a1}\markcite{a2}1989, 1990, Hernquist and Mihos
\markcite{a3}1996). More indirect scenarios are also possible, in which a
tidal galaxy interaction stimulates a burst of star formation which in turn
stimulates nonstellar nuclear activity. A further possibility is that special
conditions in isolated galaxies may trigger the feeding of fuel to an active
nucleus, such as a bar instability. (Schwartz \markcite{a31}1981; Shlosman,
Frank, \& Begelman \markcite{a2}1990; Mulchaey and Regan \markcite{a32}1997)

However, direct observational evidence that galaxy encounters stimulate the
luminosity of an AGN has been ambiguous (Adams \markcite{a4}1977, Petrosian
\markcite{a65}1983, Kennicutt and Keel \markcite{a5}1984, Dahari
\markcite{a6}\markcite{a7}1985a, 1985b, Bushouse \markcite{a8}1986,
Fuentes-Williams and Stocke \markcite{a66}1988).  One difficulty is that the
most dramatic morphological indications of the encounter may have subsided by
the time that the newly injected fuel reaches the nucleus. In any case, the
weak correlation between galaxy interactions and Seyfert activity is stronger
for type 2 Seyferts than for type 1's. 

Conversely, the presence of an AGN could alter the appearance of the central
regions of its host galaxy, principally by its injection of substantial
energy, both radiative and mechanical, over many millions of years. A further 
question is whether the particular type of active nucleus, Seyfert 1 or 2, is 
related to any property of the host galaxy. 

We have therefore used the superior spatial imaging resolution of the
post-repair {\it Hubble Space Telescope} to make a snapshot survey of nearby
active galaxies to investigate the morphological implications of different
theories on the formation and fueling of AGN. 

\section{Snapshot Survey}

In our survey program 256 images have been obtained of the cores of active
galaxies selected from the ``Catalog of Quasars and Active Nuclei" by
V\'eron-Cetty and V\'eron \markcite{50}(1987), hereafter VCV. The criteria for
choosing a galaxy from the VCV to be in our sample was that they have a
z$\leq$0.035 and not be duplicated by other cycle 4 {\it HST} observing
programs. These requirements resulted in a total of 311 galaxies. The actual
choice of the subset of 256 galaxies discussed here was random, since they
resulted from the efforts of the {\it HST} scheduling program to fill in dead
time by slewing to a nearby object for a relatively fast exposure. This
resulted in ``snapshots" of 91 galaxies with nuclear optical spectra
classified as ``Seyfert 1," 114 galaxies classified as ``Seyfert 2", and 51
galaxies classified as ``HII's." Although some of these galaxies have since
been reclassified either as intermediate Seyferts like 1.5 or 1.8, or have
been switched from Sy 1's to Sy 2's upon closer spectroscopic examination. For
our statistics we have used the most recent spectroscopic classifications from
the NASA Extragalactic Database\footnote{The NASA/IPAC Extragalactic Database
(NED) is operated by the Jet Propulsion Laboratory, California Institute of
Technology, under contract with the National Aeronautics and Space
Administration} noted in tables 1, 2, and 3. 

Some ``Active Galaxies" in the VCV are actually starbursts which are simply
included due to their very strong emission lines and are denoted as ``HII".
The line ratios are consistent with photoionization from young stars rather
than a nonstellar central engine (which emits a far larger proportion of
high-energy photons). Thus, these nuclei are radically different from Seyfert
1's, and are probably different from Seyfert 2's. We nonetheless included them
in the target list to provide a comparison with the Seyferts. 

Since the targets we have imaged constitute more than a third of all of the
nearest Seyfert galaxies currently known, they are broadly representative of
this observational class. Two biases are likely to be present because they
were present in the original searches which produced many of the entries in
the VCV. The first is that the optical discovery techniques used to find most
of these galaxies were biased against reddened, dusty active galaxies, which
are prominent in the far-infrared (\eg Spinoglio and Malkan \markcite{9}1989,
Rush \etal \markcite{10} 1993). The second bias is that the median redshift of
the Seyfert 1's (0.024) is somewhat larger than that of the Seyfert 2's
(0.017). Since the more prominent Seyfert 1 nuclei are easier to detect, they
are relatively more numerous at larger distances, where our WFPC2 images
provide a somewhat poorer linear resolution. To compensate for the effect of
the different median z's, we also did our statistical comparisons for a
modified subset of Sy 1's cut off at z=0.030, which then has a median z
close to the median z of the Sy 2's. 

The images were taken using the F606W filter because of its very high
throughput (Burrows \markcite{11}1994). This filter includes both the standard
WFPC2 V and R bands, and has a mean wavelength of 5940\AA and a FWHM of
1500\AA. We chose 500 seconds as our exposure time in a compromise between the
maximum exposure time per orbit and the minimum amount of overhead time. The
galaxy centers were usually well centered on the planetary camera CCD of
WFPC2, which has a plate scale of 0\farcs046 per pixel and a field of view of
37\arcsec\xx37\arcsec. Some of the images did not fall on the planetary camera
chip and fell on the wide field chip. Each wide field CCD has a plate scale of
0\farcs1 per pixel and a field of view of 1.3\arcmin\xx1.3\arcmin.

\section{Data Reduction.  The Atlas}

Flat field calibration, bias removal, and other initial data reduction steps
were performed at the Space Telescope Science Institute. We performed cosmic
ray removal using standard routines from the IRAF \footnote{IRAF ( Image
Reduction and Analysis Facility) is distributed by the National Optical
Astronomy Observatories, which are operated by the Association for Research in
Astronomy, Inc., under cooperative agreement with the National Science
Foundation.} software package. Unfortunately, the images have such severe
cosmic ray contamination that automated packages have difficulty in removing
all the cosmic rays. This is especially a problem for glancing hits on the CCD
which leave extended narrow trails. 

Hence the problem was to remove cosmic rays from 256 images in an efficient
manner. Doing this by hand was not practical due to the numbers involved. The
solution was to pick a threshold level which would remove most of the cosmic
ray contamination without removing any real features. This threshold level was
determined experimentally by picking multiple galaxies and running the task
{\it cosmicrays} with many different values and seeing which threshold value
eliminated the most cosmic rays without affecting any real features. The final
value that we arrived at was one where the cosmic ray would be eliminated if
its flux ratio was 80\% of the mean neighboring pixel flux. This value would
leave some residual cosmic ray contamination but would not affect real galaxy
data. We therefore chose to err on the side of caution: it is better to leave
in a cosmic ray trail rather than remove real structures. 

It must be noted here that although there is heavy contamination, the cosmic
rays are easily separable from real features within the image, since the
cosmic rays leave either a point or a line in the image. These points are
distinguishable from unresolved astronomical point sources because they do not
have the PSF surrounding the cosmic ray hit. Extended hits make linear
streaks, tightly confined to a few pixels, and thus are also easily
identifiable in contrast to real structures which have more two-dimensional
profiles. 

We did not attempt to subtract the sky background from these images, because
it was difficult to determine accurately, and relatively insignificant in any
case. The difficulty arises because in most of the images, the galaxy is more
extended than the PC chip, so that little or no true sky was measured.
Fortunately, the expected sky brightness is so faint--23rd magnitude per
square arcsecond--that it hardly effects any of our measurements or
conclusions. 

In Figures 1,2,3 we have reproduced the central 200\xx200 pixels of each image 
(with a few 400\xx400 pixel reproductions of larger galaxies), centered on the 
centroid of the galaxy nucleus.  In only a few cases (those marked with an
asterisk) is the image from one of the Wide Field chips (with a plate scale of
0\farcs1 per pixel.)  All other images are from the Planetary Camera CCD (with
a plate scale of 0\farcs046 per pixel.)  This magnification emphasizes nuclear
features that are not detectable with ground-based seeing limitations. The
grey scales are logarithmic, with full black set to the brightest pixel values
in the center of the galaxy.  In most of the Seyfert 1 images, which have
saturated nuclei, this is around 3600 Data Numbers (see Table 1).  In the
other galaxies, the brightest pixel typically has 1000 to 2000 counts.
Inevitably, a substantial amount of information is lost in this reproduction
process.

\section{Morphological Classes and Estimation of Central Magnitudes}

We have assigned morphological classes based on our images of the inner
regions of each galaxy in Tables 3, 4, and 5 on the usual Hubble tuning fork
system E/S0/Sa,Sab,Sb etc. In most cases (75\%) our classification agrees with
the one given in the Third Reference Catalog (RC3) (Corwin \etal 
\markcite{18}1994) to within one full class (\ie from Sb to Sc). Our
morphological classes for the Sy 2's are on average the same as that from the
RC3, but our Sy 1 classifications have, on average, a slight tendency to be
later than the RC3 (by less than a subclass \ie, less than the difference
between Sa and Sab). 

We have also derived azimuthally-averaged surface brightness profiles of the
centers of these galaxies to further help classify them. Fifty-two percent,
92\% and 100\% of the Seyfert 1, Seyfert 2 and non-Seyfert galaxies have
unsaturated centers. For these we have used the {\it apphot} routines in IRAF
to measure magnitudes within circular diameters of 4 pixels (0\farcs18) and 20
pixels (0\farcs92). The first aperture includes about 85\% of the light of a
point source measured by the {\it HST} Planetary Camera; the second aperture
is selected to be comparable to the seeing in good ground-based images. The
magnitude in the inner diameter, 0\farcs18, and the magnitude in the outer
diameter, 0\farcs9, get slightly dimmer with higher z, but this trend is
marginal (Figure 4). Kotilainen \etal\markcite{a62} (1993) reported 3- and
6-arcsecond aperture photometry of 5 galaxies for which we have unsaturated
images. We confirmed that there is no systematic difference between the V
magnitudes they measure and those we obtained from our data, with a scatter of
0.1 to 0.2 magnitudes. Their 3-arcsecond aperture magnitudes however, tend to
be about 0.1 magnitudes fainter than ours, which we attribute to their
ground-based seeing spilling nuclear light out of this small aperture. 

For most of the Seyfert 2 galaxies (in which there is no evidence for a
nuclear point source component), our measurements refer primarily to the bulge
light. The large-aperture magnitudes of nuclei yield a median central surface
brightness of 16.4\footnote{ Magnitudes are given in the
monochromatic-$F_{\lambda}$ Space Telescope system (Holtzman
\markcite{a54}\etal 1995). Our F606W magnitudes correspond to V magnitudes 0.1
to 0.2 magnitudes brighter. (i.e. Subtract 0.1-0.2 magnitudes from the
tabulated values to estimate V magnitudes.))}/magsq with a standard deviation
of 2.0. Byun \etal\markcite{a30} (1996) studied early type galaxies using
deconvolved PC1, F555W images. The central surface brightness magnitudes they
measured were for the area within a break radius ($r_b$) which was usually
about twice the size of our outer magnitude of r=0\farcs46. For comparison we
chose their galaxies which have relatively flat central brightness laws
($\gamma\le$0.3) so that the brightness does not rise much inside $r_b$. The
average $\mu$ for these galaxies was 16.9 $\pm$1.7, consistent with our sample
of Sy 2 galaxies. 

Phillips \etal (\markcite{a15}1996) used deconvolved PC1 images of 9
early-type disk galaxies to estimate an average surface brightness at 555nm of
16.6 /magsq in annuli from 0\farcs2 to 0\farcs5. The average surface
brightness of our Sy 2 galaxies is consistent with their value but not with
the central brightness of late-type spirals. The centers of spirals of type Sc
and later (which are disk dominated rather than bulge dominated) are nearly 2
magnitudes fainter. However our average Seyfert galaxy centers are
bulge-dominated, as expected from the very small proportion of Hubble types
later than Sc. 

For the Seyfert 1 images with saturated nuclei, we developed an indirect method
for estimating the flux from the central point source presumed to be present
in the unresolved Seyfert 1 nucleus. Although the values for radii less than a
few pixels are artificially pinned at around 3600 DN due to saturation, these
profiles are accurate and linear at radii larger than 4 to 9 pixels. 

We compared the inner portions of these galaxy profiles with those obtained
for 5 similarly bright (also saturated) stars from the same PC frames (Table
6). Their estimated total magnitudes were ``bootstrapped" by comparison with
the profiles and magnitudes of 4 field stars  with unsaturated images also
listed in Table 4. By matching the logarithmically plotted profiles over the
range of r = 4 to 9 pixels we were able to estimate the TOTAL flux from a
central point source.  Our estimates given in Table 1 are slightly too large
because we did not attempt to subtract the pedestal of galactic bulge
emission, since that correction was small and rather uncertain. Figure 5 shows
some examples of this inner-profile matching procedure.  In the good cases
(bright central point source), the uncertainty, as estimated by the scatter
between estimates from different stars, is $\pm$ 0.25 magnitudes.

\section{Measuring the Nuclear Point Source in Seyfert 1's}

Previous imaging studies have shown that nearly all Seyfert 1 nuclei emit a
featureless continuum and its ubiquitous time variability indicates that it
arises within less than a parsec of their central engine (\eg  Malkan and
Filippenko \markcite{a12}1983). Thus even with {\it HST's} resolution, this
Seyfert 1 continuum should be unresolved, and should appear as a bright point
source superposed on a resolved host galaxy. 
 
We have used our ability to discern the bright central point sources in most
of our Sy 1 sample as a method of categorizing the Sy 1's. When a nuclear
point source is evident,  we have categorized it as either a Saturated Sy 1
(SS1) or Unsaturated Sy 1 (US1). The galaxies in both these categories show a
distinct point source at their center and have a sharp rise at the 3-5 pixel
radius in their surface brightness profiles. Finally, we have grouped into the
Resolved Sy 1 (RS1) category those galaxies which were identified as Sy 1 but
showed no discernible point source, that is there was no detectable break in
their surface brightness profiles at r$\sim$3-5 pixels. If a point source is
present in these galaxies, it must typically contribute less than about 45\%
of the light within the inner one arcsecond. 
 
This classification system admittedly depends on the dynamic range of WFPC2
and the distances to the galaxies, but it has a roughly quantitative flux
basis. Out of the 91 Sy 1's 36 (40\%) fall into the SS1 category, 21 (23\%)
fall into the US1 category and 34 (37\%) fall into the RS1 category. These
ratios do not take into account the distances to the galaxies. To account for
the distances, some galaxies from the SS1 category and US1 category were
eliminated leaving a total number of 78 Sy 1's. From this sample, 30 (38\%)
fall into the SS1 category, 27 (18\%) fall into the US1 category, and 34
(44\%) fall into the RS1 category. These comparisons are summarized in table
7.

We have compared this classification system with other spectroscopic
classifications (\ie 1.8 to 1.9, Goodrich \markcite{a13}1995). Of the SS1's
only 1 galaxy is classified as a 1.8 and of the US1's only 2 galaxies have an
intermediate spectroscopic classification, but 14 galaxies in the RS1 category
have a 1.8 or 1.9 designation. Although the numbers are small, this does tend
to show that RS1's are closer to being Sy 2's than the US1's or SS1's,  in
having relatively weaker broad permitted line wings. Also those galaxies that
do not carry an intermediate spectroscopic classification may not have been
studied carefully enough to decide if such classification is appropriate. 
 
The central magnitudes of the RS1's are very similar to those of Sy 2's. The
median 0\farcs18 diameter magnitude for the Sy 2 galaxies in our sample is
19.30 with a standard deviation of 2.14. For the RS1's, the median magnitude
is 18.76 with a standard deviation of 1.23. The median 0\farcs92 diameter
magnitude of the Sy2 's is 16.86 $\pm$ 2.02, while that for the RS1's is 17.04
$\pm$ 0.93. In contrast, the US1's have a brighter 0\farcs18 diameter
magnitude at 18.53 $\pm$0.63 and a dimmer 0\farcs92 diameter magnitude of
17.13 $\pm$0.42. The dimmer outer magnitude may be accounted for by the fact
that the bulge may be dimmer in these galaxies, thus making the bright point
source relatively more prominent. 

The similarity between the Sy 2's and the RS1's also extends to their Balmer
decrements. Although very few of our observed RS1's had Balmer decrement
measurements (there are only 14), the RS1 Balmer decrements (med. =
6.13$\pm$1.51) and the Sy 2 Balmer decrements (med. = 5.34$\pm$2.53) are
higher than the Balmer decrements of the entire Sy 1 sample (med. =
4.48$\pm$2.87). The differences (at the 98\% confidence level based on the
Two-Sample Z Statistic) indicate that their nuclear continua are in many cases
weakened by dust reddening. The same does not hold true for the US1's whose
Balmer decrements (med. = 3.43$\pm$1.56) are more like those of the rest of
the Sy1's. Thus in both their spectra and their high-resolution images the
RS1's have an observational appearance similar to Sy 2's while the US1's have
an appearance and spectra similar to Sy 1's, only less powerful, and with {\it
HST's} high resolution we have been able to attach these spectral
characteristics to morphological characteristics.

\section{Unusual Seyfert 2 Galaxies}

Our atlas confirms that the nonstellar continuum is not viewed directly in
Seyfert 2 nuclei. This is the same conclusion reached by Nelson \etal 
(\markcite{a16}1996) where they observed in their WF/PC 1 images that Sy 1
galaxies tended to have strong point sources in their nuclei while Sy 2
galaxies did not. 

However, two of our classified Sy 2's do seem to show central point sources.
IC4870 is actually an ``extragalactic HII region" with unusually high
ionization lines. The saturated point source near its center is probably a
foreground star. IR1832-594 has an even brighter (saturated) central point
source. This confirms its reclassification as a Sy 1.8 by Maiolino and Rieke
\markcite{a55}(1995). 

Only 7 other Sy 2 images have saturated centers. Mrk 622 was reclassified as
an intermediate Seyfert galaxy by Goodrich \markcite{a13}(1995). Another, NGC
4507, is a heavily obscured hard X-ray (and possibly gamma-ray) source whose
X-ray properties mark it as a reddened Sy 1 (Jura \etal\markcite{a56}1997).
These bright Seyfert 2 nuclei may be intermediate cases: combinations of a
direct and a scattered Seyfert 1 continuum where the scattering region is
slightly resolved. (The center of the saturated image is often broader than a
point source). We have further evidence for its existence in another saturated
Sy 2, Mrk 533, which has broad emission lines in polarized flux. 

The remaining saturated Sy 2's (F312, F334, IR1121-281, Mrk 1370) have not
been scrutinized with such high SNR spectroscopy, which might reveal weak
broad lines. Even if they do not show polarized broad lines, there is still
another possibility that our images happened to be obtained when these Sy 2's
temporarily ``turned on" a visible Seyfert 1 nucleus (broad-line region plus
point-like nonstellar continuum). The very low frequency of saturated Sy 2
nuclei in our survey implies that such a transformation from a Sy 2 to Sy 1 is
very rare. 

\section{Quasar-like Seyfert 1 Galaxies}

A small fraction of the (more luminous) Seyfert 1 galaxies have nuclei which
are far more luminous than their host galaxies. We have singled out five of
the most extreme examples, which are indicated as ``Quasar-like" in Table 4.
In Figure 6, their azimuthally-averaged surface brightness profiles---scaled
to have matching brightness in their unsaturated wings---are overplotted (open
symbols), along with several field stars (solid symbols). At radii of less
than 3 or 4 pixels, the images are saturated to varying degrees.  At larger
radii, all of these profiles---for both stars and Quasar-like Seyfert
1's---are extremely similar. The distinctions between Seyfert 1 images and
stellar images are so subtle that they are not much larger than the
differences which appear between different stars. 

The existence of quasi-stellar Seyfert 1 galaxies is in part a selection
effect that comes from observing more luminous nuclei at higher redshifts. The
average z of these galaxies is 0.0296, and their average nuclear luminosities
are $M_{606}$= -23.6 + 5log$h_{50}$. Operationally, these Seyfert 1's can
equally well be classified as ``quasars". Three of them appear in the
Palomar-Green Bright Quasar Survey (Green, Schmidt, \& Liebert
\markcite{a52}1986).  In fact, at a slightly higher redshift they would appear
absolutely stellar in the WFPC2 images. At redshifts above a few tenths, even
more of our Seyfert 1 galaxies would appear completely unresolved in the PC2
images. These are the low-luminosity counterparts of several PG quasars imaged
by Bahcall \etal\markcite{a29}(1995) with {\it HST} which lack detectable host
galaxies (sometimes misleadingly referred to as ``naked quasars".) These
Quasar/Seyfert 1 nuclei serve as indicators that there cannot be a very good
correlation between the luminosity of the Seyfert 1 nucleus and the luminosity
of the host galaxy. 

\section{Radio Fluxes}

Integrated radio fluxes were available from the literature for 48 of our
Seyfert 1 galaxies, and 35 of our Seyfert 2's. We used these to look for a
correlation between radio luminosity and morphological classification of the
Seyfert galaxy, but found none.  The Seyfert 1 and 2 nuclei which we believe
reside in E (or E/S0) galaxies are {\it not} stronger radio emitters than
those in spirals.  This contrasts with claims that radio-loud quasars are more
likely to be found in ellipticals, while radio-quiet quasars reside in spirals
(\eg  Malkan \markcite{a61}1984). If the host galaxy morphology correlates
with radio power at high luminosities, the relation must break down at the low
luminosities of Seyfert galaxies included in the present study.

\section{Morphological Irregularities}

It is not possible to capture the rich range of morphologies seen in these
images with a simple set of classifications. Nonetheless, to draw statistical
conclusions, we have grouped most of the principal observational
characteristics we have identified into the following seven categories, most of
which are not mutually exclusive\footnote{Although no morphological
investigation of galaxy images can be completely objective, we have attempted
to note features which are clearly evident to everyone who has viewed them.
The more difficult task is in interpreting the significance of these features.
Our aim has been to provide the information, with our interpretation, but
many features may have different interpretations which are also allowed by the
data.}. 

Only $\sim$20\% of the galaxies appear, in our subjective estimation, to be
completely ``normal". By this we mean axisymmetric, with a bulge component
that has regular elliptical isophotes, and, for spirals, a thin circular
(after deprojection for inclination angle) disk that appears to be roughly
planar. Since WFPC2 is also observing non-active galaxies with unprecedentedly
high spatial resolution, it is entirely possible that many of them will now no
longer meet our definition of ``normal", either. Some subjective disagreement
is unavoidable, but our approach has been to err on the side of noting too
many possible irregularities in these images, rather than too few. 

Our search for unusual central structures is necessarily biased against
finding them near the middles of Seyfert 1 nuclei. We tried various methods of
subtracting off the bright point source; however, none were successful. Glare
from the imperfectly subtracted wings of the PSF remained, whether we used
theoretical PSF's from Tiny Tim or a library of PSF's extracted from our WFPC2
images. Thus our ability to notice low-contrast features near the galaxy
center may be greater in the normal Seyfert 2's and weaker in Seyfert 1
galaxies.  We have tested this possibility with a simulation described below.

\subsection{Evidence of Interactions}

One of our principal observational results is that few of these images show
clear indications of disturbances or other strongly asymmetric irregularities.
Two show tidal features, three have shells of differing brightness, one has
strongly warped isophotes, and 10 are in collisions/mergers. Of those galaxies
in a collision/merger, there are very few galaxies in our sample which show
two distinct nuclei in the final stages of a merger. Gorjian
\markcite{a17}(1995) has already presented 3 cases from these data of galaxies
with apparent double or triple nuclei. In only one case, Mrk 516, is this
clearly the late stage of a merger. Thus less than 0.5\% of Seyfert galaxies
show this most unambiguous evidence of being in the late stages of a merger.
If mergers are common in our sample, then one of the nuclei must not spend
very much time as a recognizably distinct sub-system when it is within the
central kiloparsec of the consuming galaxy.

\subsection{Inner Bars}

About a third of all spiral galaxies in the Third Reference Catalog  have
isophotes with strong deviations from rotational symmetry in the form of inner
isophotes with a $cos 2\Theta$ dependence, which is traditionally classified
as a strong ``Bar" (SB) or ``Lens". In the central 10 arcseconds of the PC
images, we have identified strong bars in 25 Seyfert 1's (27\%), 26 Seyfert
2's (23\%), and 6 (12\%) HII's. 

Many of these galaxies give the appearance of ``Integral Signs" or capital
``Thetas" in the centers of the galaxies. They also show up as large twists in
the isophotes as a function of radius, some of which could be equally well
described as ``Lenses". By erring on the side of caution, and only searching
the central kiloparsec regions of the galaxies covered by the PC chip, we have
probably missed many bars, especially those on larger spatial scales. This is
evidenced by bar classifications in the RC3 which do not match our
classifications. If both classifications are combined, then there are 36
(40\%) Sy 1's with bars, 58 (51\%) Sy 2's with bars, and 17 (33\%) HII
galaxies with bars. 

Barnes and Hernquist (\markcite{a19}1996) modeled barred potentials and showed
that they are effective in driving interstellar matter into the nucleus, and
fueling increased non-stellar activity there, assuming a massive black hole is
already present. We do not yet have a control sample of non-active spirals
available for comparison, but it is not apparent that strong inner bars are
unusually common in our sample of Seyfert galaxies, which is in agreement with
ground based studies (Heckman \markcite{a33}1980, Simkin, Su, \& Schwartz
\markcite{a34}1987, Mulchaey and Regan \markcite{a32}1997).

\subsection{Filaments and Wisps in Early-Type Galaxies}

Ten of the Seyfert 1's (11\%) and 24 of the Seyfert 2's (21\%) show emission
wisps or filaments which are not part of a clear spiral arm pattern (as
indicated in Tables 4 and 5 by F/W). In contrast to dust lanes, these features
are brighter than the surrounding starlight. Since the \ha/[NII] emission
lines fall near the peak sensitivity of our broad 606W bandpass, some or even
all of this extra light could be due to emission from ionized gas, which can
be confirmed by STIS spectroscopy. 

We have compared the optical morphologies of our Seyfert galaxies with
high-resolution radio maps of the same inner regions (Ulvestad and Wilson
\markcite{a20}1989, Rush \etal \markcite{a21} 1996) In most cases there is
little correlation. In some cases the filaments show some spatial correlation
with the extended radio emission, or at least some alignment, suggesting that
both are directly powered by the active nucleus. 

In spite of the large quantities of gas and ionizing photons in the H II
galaxies, they do not harbor clear examples of these filaments and wisps,
although some could have been lost in the ``noise" of the irregular background
light. We tentatively conclude that these filaments and wisps are in many
cases emission-line dominated gas associated with, and photoionized by the
active nucleus. 
 
The higher frequency of filament/wisps in the Seyfert 2's compared with the
Seyfert 1's is significant at the 96.5\% level. This result may well be
related to the fact that narrow-band ground-based imaging is more likely to
resolve the forbidden line emission in Seyfert 2's than in Seyfert 1's (Poggee
\markcite{a53}1988).

\section{Irregularities from Star Formation and Dust}

Most of the galaxies show significant deviations from smooth isophotes caused
either by localized excesses of emission (\eg star clusters and HII regions)
or localized deficits (caused by dust absorption). 

At one extreme, 5 Seyfert 1 galaxies (6\%), 11 Seyfert 2 galaxies (10\%), and
19 HII galaxies (37\%) are extremely ``clumpy." These galaxies have
large-amplitude deviations from a smooth isophotal pattern.  They contain
multiple local maxima which are at least 15\% brighter than their
surroundings. (As they are usually slightly resolved we know they are not
foreground stars in the Milky Way). In most cases these are caused by ``knots"
which are most likely active star-forming regions--star ``clusters" and their
associated H II regions. A special case of emission knots appears in the 8
galaxies we classify as ``flocculent spirals". In contrast to ``grand design"
spirals with a few very long arms, these disks have dozens to hundreds of
barely resolved patches wrapped in tight spirals. These are seen in 3\% of the
Seyfert 1's, 2\% of the Seyfert 2's, and 6\% of the HII galaxies and are
usually too fine-scale to be detected in ground-based images. 

Another special case is the 11 (12\%) Sy 1, 9 (8\%) Sy 2 and 5 (10\%) HII
galaxies with circumnuclear rings. These rings seem to be the sites of recent
star formation , and with the resolution of WFPC2 we resolve some of these
rings into many tiny ``knots."  We note that this resolving of rings into
``knots" has already been seen in the ground-based observations of NGC 7469,
by Mauder \etal\markcite{52} 1994), whose speckle masking image reconstruction
shows an excellent similarity to our PC2 image. The marked lumpiness we
observe in star-forming regions is consistent with the view that this process
may lead to the formation of large star clusters (\eg  Barth
\etal\markcite{a35} 1996) 

The H II galaxies are also more likely to appear ``Irregular" or ``Disturbed."
Again the explanation is that these galaxies have the highest fractions of
interstellar matter and associated recent star formation. In this, they
substantially exceed the average Seyfert galaxy, at least on morphological
grounds.  In other words, the host galaxies of Seyfert nuclei do {\it not}
show the extremely high star formation rates seen in the most active starburst
galaxies at the same low redshifts. 

The single most common  morphological feature evident in our images is
absorption due to interstellar dust in the active galaxy. In  many cases the
absorption has too little contrast or spatial extent to have been detectable
in ground-based images. In the absence of a standardized classification system
for dust absorption, we have attempted to identify galaxies which have dust
lanes that appear irregular. In lenticular galaxies, which we classify as E or
S0 in Tables 4 and 5, we regard  any dust absorption as noteworthy,
whereas in later-type spirals only extremely distorted dust lanes are
classified as ``irregular" (DI). 

On its face, the low fraction of Seyfert nuclei residing in dust-free and
gas-free galaxies appears to distinguish them from a randomly selected sample
of normal galaxies. This indication will remain tentative until a comparable
sample of nearby normal galaxies is examined with similar {\it HST} resolution
for signs of dust and star clusters. 

Two further cases of dust absorption of special interest have been noted even
though we do not consider them intrinsically ``irregular" for spiral galaxies.
39 galaxies appear to have dust lanes running across their nuclei, giving them
a bisected appearance (denoted DC in Tables 3,4,5) Of the Sy 2's, 23 (20\%)
were designated as DC, contrasted with 12 (13\%) of the Sy 1's. Our DC and DI
classifications are mutually exclusive. In many of these DC galaxies the small
amount of nuclear light that does reach Earth is evidently scattered back into
our line of sight, accounting for the nuclear polarizations. It also accounts
for their identification as heavily absorbed X-Ray sources: this category may
include nearly all known ``Narrow-Line X-Ray Galaxies." The nuclear dust lanes
in these galaxies have evidently obscured our direct view of the nuclear
broad-line region and optical nonstellar continuum, although the hard X-ray
emission does leak through. 

The second group of galaxies with interesting dust absorption are inclined
spirals with extensive dust lanes on one side of their major axis and hardly
any on the other side (denoted in Table 3,4,5 as D-[{\it direction}], where
{\it direction} = N, E, S, W, NE, NW, SE, SW.) 
This morphology is a simple orientation effect of the bulge light being
intercepted by the disk.  On the side of the galaxy with the disk tilted
towards us, most of the bulge light is behind, and suffers more extinction. 
The relatively smooth side is the one in which most of the bulge is in front
of the disk, which tilts away from Earth.  We have noted these cases mostly to
help break the usual indeterminacy in knowing which way the disk is tilted,
which can be of interest for comparison with the geometry of the extended
nuclear non-stellar emission (\eg in the radio). 

\section{Nuclear Dust Lanes: Intrinsic Differences Between Seyfert 1
and 2 Host Galaxies}

Forty-five (39\%) of the Seyfert 2 galaxies show either dust lanes or
absorption patches which are irregular (\ie not associated with a spiral arm
pattern) or dust that passes through the center of the galaxy  (DI and DC
categories). This is significantly higher than the proportion in Seyfert 1
galaxies, 21 (23\%) of which fall into this category. Can this difference be
attributed to the differing distances of the two samples, which might allow us
to detect more fine scale features in the closer galaxies (Sy 2's med z =
0.017), and less in the further galaxies (Sy 1's med. z = 0.024)? To test
this hypothesis, we eliminated the Sy 1 and Sy 2 galaxies that had a
z$\geq$0.03 from our sample thereby bringing the median z of the two samples 
closer together (Table 8). For both Sy 1 and Sy 2 galaxies, the percentage of 
DC and DI galaxies went up, but the difference in dustiness between the two 
categories changed very little. Before the z based selection the difference 
was 16\%, after the z based selection it was 14\%. Thus we are not likely 
missing large numbers of DC and DI Sy 1's because of their greater distances.

Another selection effect which may cause the Sy 1's to seem less dusty than
the Sy 2's would be because of glare from saturated point sources. We have
tested the possibility that dust closer to the nucleus is lost when a
saturated point source is present with a simulation.  One of us added observed
saturated point sources to an anonymous subsample of representative Seyfert 2
images, thus converting them into artificial SS1's. These test images were then
reclassified by another one of us, who {\it independently} obtained the same
dust absorption classification in 84\% of the galaxies as in their original
images. 

In only one case-- ESO 373-G29 --was dust absorption missed when the saturated
point source was added, because it is only visible very close to the center of
that galaxy. The only other misclassification went in the opposite sense: IR
2246-195 was noted as having a dust lane which ``possibly" extended close to
the center (\ie, with less confidence than any of the other dust detections in
the test images). This test indicates that the deficiency of dust detections
in the SS1's relative to the Seyfert 2's is not due to the glare of the point
source.  Such an explanation would require that we missed {\it half} of the
dust features in the SS1's, whereas we in fact missed only one-sixth of them.

Furthermore this effect cannot explain the low rate of dust detections in the
Unsaturated Seyfert 1's, which is still significantly below that of the
Seyfert 2's. The DC/DI percentage in the Sy 1's with point sources (SS1 and
US1 category) is 19\% (11 out of 57) versus the rate for Sy 1's with no point
sources (RS1 category) which is 29\% (10 out of 34). This is still below the
39\% for the Sy 2's. 

This difference is also seen in the Hubble types we assigned for the inner
region of each galaxy.  Figure 7 shows that the Seyfert 1 galaxies are more
skewed to Sa types, and away from Sc's. (The ratio of Sa/Sc galaxies with
Seyfert 1 nuclei is 3.9; it is 1.0 for Seyfert 2 galaxies). This skewness
toward early-types in Seyfert 1's also shows up in the median morphological
class, which is Sa for the Seyfert 1 galaxies and Sb for the Seyfert 2's. 

The combination of these robust differences in morphological classes and
dust-lane classifications indicate that the centers of Seyfert 2's are {\it
intrinsically more dusty} than the centers of Seyfert 1's. Our spatial
resolution is many orders-of-magnitude coarser than the structures that define
a Seyfert 1--the broad-emission-line region and the even more compact
nonstellar continuum. Thus we could easily fail to see a small opaque dust
cloud which is nonetheless large enough to block out the Seyfert 1 nucleus in
many galaxies we classify as Seyfert 2's. It seems likely, however, that the
dust we see on larger scales would be statistically associated with these
small (possibly unseen)  dust clouds. Therefore, in a dustier galaxy, more
lines-of-sight to the central active nucleus are likely to intersect small but
opaque galactic dust clouds. We postulate that these galactic dust clouds are
a major reason we cannot see the Seyfert 1 nucleus directly in Seyfert 2
galaxies. 

Our imaging evidence for galactic dust clouds along our lines-of-sight to
Seyfert 2's is statistical, since we would not be able to see them in every
case. The absorption must have a great deal of unresolved fine structure.
Since {\it HST} cannot resolve the small spatial scales on which the Seyfert 1
nucleus can be blocked, we could hardly expect a one-to-one correspondence
between a dusty appearance and a Seyfert 2 classification. Thus we have
probably missed small galactic dust lanes which lie in front of some Seyfert 2
nuclei which did not appear ``dusty" in our PC images. Conversely, it is also
plausible that those ``dusty" Seyferts classified as type 1 happen to have a
relatively dust-free gap along our line-of-sight, which allows at least some
of the nuclear light to reach us directly. {\it HST}'s order-of-magnitude
improvement in spatial resolution may well have shown us the tip of the
iceberg. With another comparable improvement in resolution, we predict that
the correlation between nuclear dust absorption and Seyfert 2 classification
will become even stronger than it is on our data. 

The suggestion that Seyfert 2 nuclei are more heavily obscured is not new (\eg
Lawrence and Elvis \markcite{a22}1982, Malkan and Oke \markcite{a23}1983).
They are redder than Seyfert 1's at all wavelengths from the far-infrared to
the X-rays, and a relatively larger fraction of their total energy output has
been reprocessed by warm dust grains  (Edelson and Malkan \markcite{a24}1986).
Because the surviving transmitted continuum is weaker, the scattered continuum
light becomes relatively more prominent. The greater dust covering fraction
enhances the scattering, and increases the proportion of nuclear luminosity
which is re-radiated in the thermal infrared (Spinoglio and Malkan
\markcite{a9}1989). 

We noted above that our full sample may suffer some selection biases. The
principal one is that Seyfert 2's are less likely to be included unless they
are relatively prominent (\ie with unusually high nuclear luminosities or
star formation rates). This raises the danger that we are comparing the
Seyfert 1 galaxies with Seyfert 2's which are intrinsically more luminous. We
have no reason to suppose that more luminous Seyfert 2's should have higher
dust covering fractions than less luminous Seyfert 2's. Nonetheless, in
testing predictions of the unified scheme, it is desirable to compare Seyfert
1s and 2s which are matched in some isotropic emission property. We have
compared those Seyferts with measured low-frequency radio fluxes and [OIII]
5007 emission line fluxes. These quantities, which are believed to be
relatively orientation-independent, have {\it indistinguishable distributions in
our Sy 1's and Sy 2's}. Thus we have avoided the most common pitfall in
constructing Sy 1/Sy 2 comparison samples to test unified schemes. 

Another test is to consider the subsample of Seyfert galaxies which were not
selected in the traditional optical/UV searches. Fortunately 29 of our Sy 1's
and 31 of our Sy 2's are members of the 12 Micron Galaxy Sample which avoids
the usual biases of these other methods (Spinoglio and Malkan
\markcite{a9}1989; Rush \etal \markcite{a10}1993). The morphologies of this
subset show the same preference for Sy 1 nuclei to reside in earlier-type
galaxies than Sy2 nuclei (Figure 8). The median morphological class is Sab for
the Sy 1's and Sbc for the Sy 2's. Since this subsample shows the same
distinction as our entire (heterogenous) sample, we believe that the effect is
not an artifact of selection. 

\section{Seyfert Unification: An Alternative to the Orientation Hypothesis}

The Seyfert unification hypothesis states that each Seyfert 2 nucleus actually
harbors a normal Seyfert 1 nucleus in its center (Peterson
\markcite{a36}1997). It has long been suspected that the classical
observational signatures of the Seyfert 1 nucleus (point-like nonstellar
continuum plus broad permitted emission lines) are not visible in Seyfert 2's
because of obscuration along our line-of-sight to the central engine (\eg
Lawrence and Elvis \markcite{22}1982; Malkan and Oke \markcite{a23}1983). All
of the Seyfert 1 nuclei which suffer sufficient extinction (corresponding to
more than several magnitudes of visual absorption) will then appear to us as
Seyfert 2's. This idea is well established for some Seyfert 2 nuclei which
show broad emission lines in polarized light. This powerful observational
signature has by no means been shown to be universal among Seyfert 2's.
Whether {\it all} Seyfert 2's harbor obscured Seyfert 1 nuclei is still
controversial. (See Lawrence\markcite{a25}1991 for a review). The present
imaging study does not settle this question. In the following discussion we
will assume the unification hypothesis is true, and use our data to make
further inferences.

The differences in the apparent nuclear dustiness of the Seyfert 1 and 2 
host galaxies in these PC images should not depend much, if at all, on the
orientation of the nucleus (or the orientation of the galaxy, if that were
similar). Our data suggest that this difference does not result from viewing
angle, but because {\it a greater fraction of the sky as seen from a Seyfert 2
nucleus is blocked by obscuration.} This is at least a complication to the
simple ``unified scheme" in which Seyfert 2's are intrinsically identical to
Seyfert 1's except for the angle at which they are viewed. Instead, we argue
that those Seyferts which are classified as type 2 are more likely to have
larger ``dust covering fractions" than the average type 1's.  This {\it
intrinsic difference} could explain some of their different observed broadband
properties (\eg  Edelson \etal \markcite{a26}1987, Carleton \etal
\markcite{a27} 1987), as well as most of the usual difficulties with the
simple unified scheme (\eg review by Antonucci \markcite{a28} 1993).

Still assuming the Seyfert unification hypothesis, an outstanding question
remains: What is the nature and location of the absorbers that obscure our
view of the Seyfert 1 nucleus, which we suppose is always present in the
center of a Seyfert 2 galaxy? One possibility has become well known since it
was sketched in Antonucci (\markcite{a37}1982). That paper described a
particular scenario in which our view of the central continuum and broad-line
region is occulted by a dusty thick gas ring which surrounds the central
engine, and is closely aligned with its rotation axis. A Seyfert which we view
at sufficiently high inclination, through its torus, will appear to us as a
type 2. A Seyfert nucleus which we view close enough to a face-on orientation
would show us a direct view of its BLR, and would thus be classified as type
1. This Accreting Torus Model (ATM) is sketched on the left-side panel of
Figure 9. 

This specificity also makes the AT model vulnerable to observational tests.
Based on the observed fact that Seyfert 2's are somewhat more numerous locally
than Seyfert 1's requires that the opening angle of the torus cannot be larger
than about $\pi$ steradians (Edelson \etal \markcite{a26}1987; Rush \etal
\markcite {a10}1993). The inner diameter need only be large enough to engulf
the BLR, hardly more than a parsec for the typical low-luminosity Seyferts in
our survey. The outer diameter is not expected to be much more than one or two
orders of magnitude larger, so that the entire torus structure remains aligned
with (and connected to) the central engine, rather than the host galaxy. More
than 100 parsecs from the central engine, its gravity is likely to be less
important than that of the galactic stars. Assuming the obscuring torus is a
small extension of the central engine, and is relatively independent of the
host galaxy, we formulated two expectations about our WFPC2 images: a) the
obscuring torus will be too small to detect at our typical resolution of a few
hundred parsecs; and b) it will not be connected with the galactic dust lanes
which we can observe hundreds of parsecs away from the galactic center; and c)
therefore any difference between Seyfert 1 and 2 nuclei would only be invisible
on the larger size scales probed by our direct images. 

These admittedly simplistic expectations are not borne out by our
observations. The higher observed incidence of irregular dust absorption in
the centers of Seyfert 2 galaxies suggests that we are in some cases directly
observing the source of the nuclear extinction: interstellar dust clouds which
intercept our line-of-sight to the nucleus. These dust lanes are seen on
scales of hundreds of parsecs, and may therefore have {\it little or no
physical connection with the central engine}. This alternative to the AT
model, the Galactic Dust model (GDM), is illustrated schematically on the
right-hand panel of Figure 9. 

The viability of the GDM depends on whether galactic dust outside the nucleus
could produce enough extinction to transform a Seyfert 1 nucleus into a
Seyfert 2? The answer depends on how much extinction is required, and how much
might plausibly be available on scales of a hundred parsecs. The visible and
UV traces of a Seyfert 1 nucleus (broad lines and compact continuum) would be
substantially obliterated at an extinction above $A_V$ of 10 magnitudes. Even
unusually deep infrared searches are hard pressed to detect buried Seyfert 1
nuclei when $A_V$ is 25 to 50 magnitudes (Ward \etal\markcite{a68}1991).
These extinctions would completely block soft X-ray emission from the nucleus.
Thus for a normal dust/gas ratio the line of sight to the center of a typical
Seyfert 2 galaxy would need to intercept 0.2--1.0 \xx 10$^{23}$ atoms
cm$^{-2}$. Such column densities are roughly consistent with the average values
inferred in Seyfert 2's from limited hard X-ray spectroscopy (Mulchaey \etal\
markcite{a62}1992). 

This amount of extinction would be produced if our line-of-sight to the
galactic nucleus intercepted about 10 diffuse molecular clouds, or a single
dense molecular cloud core. This might happen in a Milky-Way-type disk at a
radius of $\sim$100 parsecs, where the vertical extent of the molecular gas
(including extra-planar warps), is comparable to the distance from the nucleus
(Sanders, Solomon, \& Scoville\markcite{a69}1984). In fact, the average
surface density of molecular gas in the inner 500 parsecs of the Milky Way
corresponds to N$_H$ = 4 \xx 10$^{22}$ atoms cm$^{-2}$.  This value is
probably typical (it is, for example, 10$^{23}$ in the inner few hundred parsecs
of Maffei 2--Hurt\markcite{a70} 1994). Furthermore, our images indicate that
the interstellar dust is often {\it disturbed} in Sy 2's, so that it is not
confined to a 100-pc-thick slab in the disk plane. 

The ATM is most likely to be applicable to those  Seyfert 2's which have Fe
K$\alpha$ emission lines of enormous equivalent width.  Assuming this Fe line
is produced by X-ray fluorescence, it implies the existence of a bright hard
X-ray continuum which is obscured by N$_H$ $\ge$ 10$^{24}$ cm$^{-2}$, which
might be too large for the GDM to explain. There are only a handful of such
Seyfert galaxies known currently, NGC 1068 being the most famous (Smith, Done,
\& Pounds\markcite{a67}1993), but few sufficiently sensitive observations have
been made.

If the GDM is more applicable to most Seyfert nuclei than the ATM, then
several major implications would follow: 
 
1) The obscuring region in Seyfert 2's need not surround it on all sides.
Complete 360-degree azimuthal symmetry cannot be assumed, and there need not be
any well defined ``opening angle".

2) Galaxy interactions may be an important mechanism for (temporarily)
increasing the dust covering fraction of an active galactic nucleus. 
According to the GDM, this active nucleus is then more likely to
appear as a Seyfert 2. This would be consistent with a higher rate
of galaxy interactions in Seyfert 2's than in Seyfert 1 galaxies.

3) The obscuring region may be typically more than 100 parsecs from the
central engine.  The mutual physical influence of these two regions on each
other may be very small. 

4) The obscuring region could be observed in emission, but only at
long wavelengths. For a typical nuclear luminosity of 10$^{43}$ erg/sec,
dust grains at a radius of 100 parsecs should reach an equilibrium 
temperature around 50 K.  Thus if Sy 2's do have dustier centers than
Sy 1's, they could be relatively stronger emitters of far-infrared
continuum and perhaps CO line emission. Testing this prediction would,
however, require observations at wavelengths longer than 100\mm.
Unfortunately, the spatial resolution required to isolate and map the central
100 parsecs is of order an arcsecond, probably achievable via interferometry.

5) The orientation of this dust lane may have little or no relation to the
{\it intrinsic} orientation of the central engine. Of course it does affect
what we are able to see from Earth. In particular, it blocks our direct view
of the central continuum source and the broad line region. It also
extinguishes scattered photons from these regions unless they are escaping out
of the plane of the dust lane.  This explains why some Seyfert 2 nuclei have
linear continuum polarizations which are modest (typically a few up to 15\%),
but always less than would be expected for the very simple, well-defined
scattering geometry of a torus plus conical-funnel scattering region (where
the expected polarizations can range from 20\% to much higher values). 

6) The linear or even bi-polar structures which are sometimes seen in the
[OIII] line and radio continuum emission in Seyfert nuclei do not require the
presence of a geometrically thick dust torus for their existence. These
structures, which are sometimes spatially related, appear to be produced by a
bi-polar outflow of energy along the two opposite poles of the central engine.
The central engine tends to eject energy (which could be in the form of
relativistic particles or mechanical energy) along what is thought to be the
spin axis. As shown in Figure 9, in the ATM, only gas above the poles of the
torus sees the ionizing radiation from the central source. In the GDM the
central ionizing continuum and the [OIII] line emission need not be bipolar,
except for the component of the NLR which is associated with the radio jet. In
the few cases where dense molecular gas has been detected orbiting close to an
active nucleus, such as NGC 4258, it lies in an extremely thin disk, not a
torus. 

7) Even though the central engines in Seyfert 1 and 2 nuclei may be
intrinsically the same, the inner regions of their host galaxies are {\it
not}. If Seyfert 2 galaxies have nuclei which are covered by a
larger fraction of dust clouds, they would be more often observed 
as Seyfert 2's rather than less obscured Seyfert 1's. This larger areal
dust covering fraction is likely to have additional observational
manifestations, including more heavily reddened emission line ratios and
different thermal dust emission spectra in the infrared (\eg Edelson \etal
\markcite{a26}1987). 

8) If much of the obscuring medium is not fixed to the central engine, it may
well be moving across our line-of-sight. Typical orbital speeds for galactic
dust clouds are up to 0.1\% the speed of light. Thus fine structure in the
dust lane can traverse the nucleus in as little as 1000 light-crossing times. 
This would correspond to changes in the reddening/extinction of the nuclear
continuum on timescales as short as weeks, and changes in the
reddening/extinction of the broad emission lines over less than a decade.
Statistically, we could expect the partially covered nuclei (our RS1's) to
show the strongest reddening variations, and much more so in the continuum
than in the lines.  If the projected cloud edges are relatively straight as
they move across the nucleus, this kind of extinction variability could
provide a kind of tomography of the central engine, somewhat different from
the information gleaned from reverberation. 

9) Assuming the central engine of the Seyfert nucleus has a symmetry
(rotation?) axis, its viewing angle is probably {\it not} simply related to its
apparent classification. Thus, for example, we probably view some Seyfert 1
nuclei at high inclinations, just as we probably view some Seyfert 2 nuclei
close to ``face-on." 

\section{Summary}

Our large sample of high-resolution images of the centers of nearby Seyfert 1,
2 and HII galaxies has allowed us to search for statistical differences in
their morphologies. 

The Seyfert galaxies do not, on average, resemble the HII galaxies. The latter
have more irregularity and lumpiness associated with their high rates of
current star formation. Conversely, none of the HII galaxies have the
filaments or wisps which are sometimes seen in Seyfert 1 and 2 galaxies, and
are evidently gas filaments photoionized by the active nucleus. 

Sixty-three percent (63\%) of the galaxies classified as Seyfert 1 have an
unresolved nucleus, 52\% of which are saturated. Some (6\%) have such dominant
nuclei that they would appear as ``naked quasars" if viewed at somewhat higher
redshifts. The presence of an unresolved nucleus, particularly a saturated
one, is anti-correlated with an intermediate spectroscopic classification
(such as Seyfert 1.8 or 1.9) and is also anti-correlated with the Balmer
decrement.  This implies that those Seyfert 1's with weak nuclei in the PC2
images are extinguished and reddened by dust. 

The vast majority of the Seyfert 2 galaxies show no central point source. In
fact, the only two of these that do (IRAS 1832-594 and IC 4870) are
mis-classified galaxies. If all Seyfert 2's actually harbor point-like
continuum sources like those in Seyfert 1's, they are at least an order of
magnitude fainter on average. In those galaxies without any detectable central
point source (37\% of the Seyfert 1's; 98\% of the Seyfert 2's, and 100\% of
the H II's), the central surface brightnesses are statistically similar to
those observed in the bulges of normal galaxies. 

Seyfert 1's and 2's  both  show circumnuclear rings in about 10\% of the
galaxies. We identified strong inner bars as often in Seyfert 1 galaxies
(27\%) as in Seyfert 2 galaxies (22\%). In some cases we see a strong
assymetry of the dust absorption across the major axis, which allows us to
infer which half of the disk is nearer to us: the side which more strongly
absorbs the smooth light of the bulge behind it. 

The Seyfert 2 galaxies are more likely than Seyfert 1's to show irregular or
disturbed dust absorption in their centers as well as galactic dust lanes
which pass very near their nuclei. They also, on average, tend to have latter
morphological types than the Seyfert 1's. This difference remains in Seyfert 1
and 2 subsamples matched for redshift, [OIII] and radio luminosities.  It also
holds true when we restrict our consideration to sub-samples of the data which
are less biased by selection effects.  Thus it appears that the host galaxies
of Seyfert 1 and 2 nuclei are {\it not} intrinsically identical. A galaxy with
more nuclear dust and in particular more irregularly distributed dust is more
likely to harbor a Seyfert 2 nucleus. This indicates that the higher
dust-covering fractions in Seyfert 2's are the reason for their spectroscopic
classification: their compact Seyfert 1 nucleus may have been obscured by
galactic dust. This statistical result contradicts the simplest and most
popular version of the unified scheme for Seyfert galaxies. We suggest that
the obscuration which converts an intrinsic Seyfert 1 nucleus into an apparent
Seyfert 2 often occurs in the host galaxy hundreds of parsecs from the
nucleus. If so, this obscuration need have no relation to a hypothetical fat
dust torus surrounding the equator of the central engine. Also then the
orientation of the central engine with respect to our line-of-sight does {\it
not} determine whether an active nucleus will appear to us as a Seyfert 1 or
as a Seyfert 2.

\acknowledgements

We thank Wayne Webb and Randall Rojas for help in the early stages of this
research, and M. Regan for insightful refereeing. This research has made use
of the NASA/IPAC Extragalactic Database (NED), which is operated by the Jet
Propulsion Laboratory, Caltech, under contract with NASA.

\newpage 

\centerline{\bf FIGURE CAPTIONS} 

\noindent {\bf Figure 1} --- These are images of galaxies that in NED were
classified as Seyfert 1 - 1.9. The arrowhead points North and the bar is East.
The length of the eastern bar is 2" in the PC images. Those galaxies that fell
on the Wide field chip are designated with an asterisk. The length of the
eastern bar in the WF images is 4". The scale is based on $H_o$=50 \ksm. Note
the strong point source in the centers of these galaxies and the fact that the
host galaxies are generally earlier in type. 

\noindent {\bf Figure 2} --- These are images of galaxies that in NED were
classified as Seyfert 2's. The arrowhead points North and the bar is East. The
length of the eastern bar is 2" in the PC images. Those galaxies that fell
on the Wide field chip are designated with an asterisk. The length of the
eastern bar in the WF images is 4". The scale is based on $H_o$=50
\ksm. Note that the host galaxies appear to be later in type and that some
of their dust is in a non-spiral pattern. 

\noindent {\bf Figure 3} --- These are images of the galaxies that in NED
were classified as HII region galaxies. The arrowhead points North and the bar
is East. The length of the eastern bar is 2" in the PC images. Those
galaxies that fell on the Wide field chip are designated with an asterisk. The
length of the eastern bar in the WF images is 4". The scale is based on
$H_o$=50 \ksm. NGC's 625, 3738, 4700, 5253 were too large so they have
been reproduced more extensively. Note that although most of the galaxies are
very complex, they do not show any emission filaments or wisps. 

\noindent {\bf Figure 4} --- These are plots of redshift vs. inner and outer 
magnitude. The upper plot is for the magnitudes with a radius of 0\farcs09 and
the lower plot is for the magnitudes with a radius of 0\farcs46. The filled
symbols are from our derived magnitudes of the saturated centers. 

\noindent {\bf Figure 5} --- These are examples of matching the brightness
profile of a point source to the central brightness profile of a galaxy. The
cross symbols represent the profile of the galaxy and the star symbols
represent the profile of the star. The profile of the star is multiplicatively
scaled to match the profile of the galaxy at the 3-5 pixel range. 

\noindent {\bf Figure 6} --- These are radial profiles of a sample of stars, 
filled symbols, and our 5 quasar-like Sy 1's (open symbols). Note the 
similarity of the two profiles and how the differences between the Sy 1's is 
no more than the differences between the different stars.

\noindent {\bf Figure 7} --- These are histograms of the Hubble classes for
the full sample of Sy 1 and Sy 2 galaxies. The black part of the histograms
represent the number of barred galaxies in each class. Note the higher
proportion of Sy 1's in Sa's relative to Sc 's. 

\noindent {\bf Figure 8} --- These are histograms of the Hubble classes for
the 12\mm sample of Sy 1 and Sy 2 galaxies. The black part of the histograms
represent the number of barred galaxies in each class. Note that in this
different sub-sample that a higher proportion of Sy 1's in Sa's relative to Sc
's. 

\noindent {\bf Figure 9} --- Two schematic representations of competing
schemes for unifying Sy 1's and Sy 2's. The Accreting Torus Model (ATM)
requires a geometrically thick 
dusty torus to block out most of the light from the central
Sy 1 engine, while the Galactic Dust Model (GDM) depends on dust obscuration
present in the inner regions of the host galaxy. In the GDM model the Narrow
Line Region (NLR) has two energizing mechanisms. The accretion disk provides a
more distributed energy output, while jets from the central black hole would
provide energy for `` ionization cones." Due to space limitations, the counter
jet/cones have not been included in the schematic. In the ATM, Sy 1's are
viewed closer to pole-on, while Sy 2's are viewed closer to edge-on.  In the
GDM this generalization does not hold.

\end{document}

\begin{deluxetable}{lllccc}
\footnotesize
\tablenum{2}
\tablewidth{0pt}
\tablecaption{Seyfert 2 Galaxies}
\tablehead
{
\colhead{Galaxy} &
\colhead{Alternate} &
\colhead{z} &
\colhead{12\micron} &
\colhead{Inner} &
\colhead{Outer}
\\[.2ex]
\colhead{} &
\colhead{Designation} &
\colhead{} &
\colhead{} &
\colhead{Mag.\tablenotemark{a}} &
\colhead{Mag.\tablenotemark{b}}
}

\startdata

ESO 137-g34      &                 & 0.009  &        &   19.28  &   15.90    \nl
ESO 138-g01      &                 & 0.009  &        &   16.69  &   14.37    \nl
ESO 139-g12      &                 & 0.017  &        &   16.87  &   14.83    \nl
ESO 353-g9       &                 & 0.017  &        &   19.25  &   16.45    \nl
ESO 362-g8       &                 & 0.016  &        &   16.41  &   13.50    \nl
ESO 373-g29      & IC 2510          & 0.009  &        &   18.89  &   16.93    \nl
ESO 509-g66      &                 & 0.034  &        &   18.54  &   16.42    \nl
ESO 509-g66-c    &                 & 0.045  &        &   20.64  &   18.52    \nl
FRL 294            &                 & 0.017  &        &   19.21  &   16.83    \nl
FRL 312            & IC 3639          & 0.011  &        &   [17.06]  &  \nodata   \nl
FRL 315            &                 & 0.016  &        &   18.90  &   16.47    \nl
FRL 316            &                 & 0.016  &        &   19.95  &   17.36    \nl
FRL 334            & IC 4777          & 0.018  &        &   [17.58]  &  \nodata   \nl
FRL 341            & IC 4995          & 0.016  &        &   19.03  &   16.68    \nl
IC 184           &                 & 0.018  &        &   20.96  &   18.03    \nl
IC 4870          &                 & 0.003  &        &   [19.54]  &  \nodata   \nl
IRAS 0147-076      & IRAS 01475-0740    & 0.017  & X      &   19.24  &   17.56    \nl
IRAS 0258-1136     & MCG -2-8-39      & 0.030  & X      &   17.84  &   17.20    \nl
IRAS 0450+039      & CGCG 420-015     & 0.030  &        &   18.84  &   16.99    \nl
IRAS 0450-032      & PGC 16226        & 0.016  &        &   20.88  &   18.28    \nl
IRAS 0457-756      & ESO 33-g2        & 0.019  & X      &   19.23  &   16.80    \nl
IRAS 1121-281      & IRAS 11215-2806    & 0.014  &        &   -----  &   -----    \nl
IRAS 1305-241      & IRAS 13059-2407    & 0.014  &        &   19.81  &   16.84    \nl
IRAS 1443+272      & IRAS 14434+2714    & 0.029  &        &   19.11  &   17.43    \nl
IRAS 1548-037      & IRAS 15480-0344    & 0.030  & X      &   19.29  &   17.12    \nl
IRAS 1832-594      & FRL 49             & 0.019  &        &   [17.58]  &  \nodata   \nl
IRAS 1833-654      & ESO 103-g35      & 0.013  &        &   19.34  &   16.86    \nl
IRAS 2246-195      & MCG -3-58-7      & 0.033  & X      &   18.83  &   16.55    \nl
IRAS 2302-000      & UGC 12348        & 0.030  &        &   19.35  &   16.95    \nl
IRAS 2346+019      & IRAS 23461+0157    & 0.031  & X      &   19.11  &   17.82    \nl
MCG -5-27-13     & ESO 439-g9       & 0.023  &        &   21.31  &   18.25    \nl
MRK 1           &                 & 0.016  &        &   18.45  &   16.78    \nl
MRK 3           & UGC 3426         & 0.014  &        &   18.39  &   15.59    \nl
MRK 176         & UGC 6527         & 0.027  &        &   18.87  &   16.53    \nl
MRK 198         &                 & 0.024  &        &   19.71  &   17.00    \nl
MRK 266sw       & NGC 5256           & 0.028  & X      &   20.77  &   18.00    \nl
MRK 270         & NGC 5283           & 0.009  &        &   18.74  &   16.17    \nl
MRK 313         & NGC 7465           & 0.006  &        &   18.99  &   16.40    \nl
MRK 348         & NGC 262            & 0.014  & X      &   18.82  &   16.77    \nl
MRK 403         &                 & 0.024  &        &   20.14  &   17.87    \nl
MRK 533         & NGC 7674           & 0.029  & X      &   [17.58]  &  \nodata   \nl
MRK 573         &                 & 0.017  &        &   18.97  &   16.37    \nl
MRK 577         & UGC 1282         & 0.017  &        &   19.30  &   16.63    \nl
MRK 607         & NGC 1320           & 0.009  & X      &   18.81  &   16.59    \nl
MRK 612         &                 & 0.020  &        &   19.61  &   16.90    \nl
MRK 620         & NGC 2273           & 0.006  &        &   19.24  &   16.64    \nl
MRK 622         & UGC 4229         & 0.023  &        &   [17.30]  &  \nodata   \nl
MRK 686         & NGC 5695           & 0.014  &        &   19.51  &   16.77    \nl
MRK 917         &                 & 0.025  &        &   19.05  &   16.58    \nl
MRK 937         &                 & 0.030  &        &   19.41  &   17.68    \nl
MRK 938         & NGC 34             & 0.019  & X      &   18.78  &   16.31    \nl
MRK 955         &                 & 0.035  &        &   20.49  &   17.58    \nl
MRK 993         & UGC 987          & 0.017  &        &   19.28  &   16.85    \nl
MRK 1058        &                 & 0.018  &        &   19.35  &   17.09    \nl
MRK 1066        & UGC 2456         & 0.012  &        &   18.67  &   16.24    \nl
MRK 1073        &                 & 0.023  &        &   19.63  &   16.84    \nl
MRK 1157        & NGC 591            & 0.015  &        &   19.20  &   16.86    \nl
MRK 1193        &                 & 0.032  &        &   19.74  &   17.57    \nl
MRK 1210        & UGC 4203         & 0.013  &        &   18.40  &   16.46    \nl
MRK 1370        &                 & 0.024  &        &   18.91  &   17.16    \nl
NGC 424            & TOL 0109-383     & 0.011  & X      &   [18.05]  &  \nodata   \nl
NGC 513            &                 & 0.016  & X      &   20.08  &   17.33    \nl
NGC 788            &                 & 0.013  &        &   18.95  &   16.52    \nl
NGC 1125           &                 & 0.011  & X      &   17.73  &   17.25    \nl
NGC 1144           &                 & 0.029  & X      &   20.29  &   17.33    \nl
NGC 1241           &                 & 0.013  & X      &   19.90  &   17.22    \nl
NGC 1358           &                 & 0.013  &        &   19.58  &   16.62    \nl
NGC 1386           &                 & 0.002  & X      &   18.92  &   15.86    \nl
NGC 1410           & IIIZW55N        & 0.025  &        &   20.50  &   17.48    \nl
NGC 1667           &                 & 0.015  & X      &   19.81  &   16.93    \nl
NGC 2110           &                 & 0.007  &        &   18.86  &   16.50    \nl
NGC 2992           &                 & 0.007  & X      &   19.22  &   16.91    \nl
NGC 3081           &                 & 0.007  &        &   19.19  &   16.82    \nl
NGC 3362           &                 & 0.028  &        &   20.38  &   17.96    \nl
NGC 3393           &                 & 0.012  &        &   19.49  &   16.57    \nl
NGC 3982           &                 & 0.003  & X      &   19.29  &   17.12    \nl
NGC 4156           &                 & 0.022  & X      &   19.33  &   17.34    \nl
NGC 4507           & ESO 322-g29      & 0.012  &        &   [17.05]  &  \nodata   \nl
NGC 4922b          &                 & 0.024  & X      &   19.34  &   16.71    \nl
NGC 4939           &                 & 0.010  &        &   19.84  &   16.84    \nl
NGC 4968           & ESO 508-g6       & 0.009  & X      &   18.82  &   16.76    \nl
NGC 5135           & ESO 444-g32      & 0.013  & X      &   19.13  &   16.81    \nl
NGC 5347           &                 & 0.008  & X      &   19.10  &   16.83    \nl
NGC 5427           &                 & 0.009  &        &   19.72  &   17.52    \nl
NGC 5929           & UGC 9851         & 0.008  & X      &   20.00  &   17.03    \nl
NGC 5953           &                 & 0.007  & X      &   19.05  &   16.00    \nl
NGC 5995           &                 & 0.025  & X      &   19.00  &   16.82    \nl
NGC 6211           &                 & 0.020  &        &   18.25  &   15.55    \nl
NGC 6217           &                 & 0.005  & X      &   18.32  &   15.63    \nl
NGC 6221           & ESO 138-g3       & 0.004  &        &   19.56  &   16.55    \nl
NGC 6300           &                 & 0.003  &        &   20.59  &   17.63    \nl
NGC 6393           & IRAS 17296+5940    & 0.028  &        &   21.92  &   19.05    \nl
NGC 7130           & IC 5135          & 0.016  & X      &   18.53  &   16.02    \nl
NGC 7172           &                 & 0.008  & X      &   21.63  &   18.14    \nl
NGC 7212           & UGC 11910        & 0.026  &        &   19.41  &   16.94    \nl
NGC 7319           & UGC 12102        & 0.022  &        &   20.95  &   18.16    \nl
NGC 7410           &                 & 0.006  &        &   20.57  &   17.33    \nl
NGC 7582           & ESO 291-g16      & 0.005  & X      &   19.27  &   16.43    \nl
NGC 7590           &                 & 0.005  & X      &   19.82  &   16.87    \nl
NGC 7592           &                 & 0.024  &        &   20.87  &   18.44    \nl
NGC 7682           & UGC 12622        & 0.017  &        &   20.16  &   17.36    \nl
NGC 7743           &                 & 0.007  &        &   18.30  &   15.54    \nl
PKS 2048-572     & IC 5063          & 0.011  & X      &   19.46  &   17.10    \nl
PKS 2158-380     &                 & 0.033  &        &   19.50  &   16.79    \nl
Q 1234+0848      & TOL 1234+088     & 0.028  &        &   21.69  &   19.74    \nl
UGC 3255         &                 & 0.019  &        &   21.68  &   18.77    \nl
UGC 4332         & A 0816+21        & 0.018  &        &   21.04  &   17.78    \nl
UGC 6100         & IRAS 10587+4555    & 0.029  &        &   19.23  &   17.02    \nl
UM 105           & IC 123           & 0.030  &        &   20.83  &   18.18    \nl
UM 319           &                 & 0.016  &        &   21.66  &   18.65    \nl
UM 625           &                 & 0.025  &        &   19.55  &   17.55    \nl
WAS 2            &                 & 0.033  &        &   19.01  &   16.67    \nl
ZW 1408+137      & IRAS 14082+1347    & 0.017  &        &   19.88  &   17.71    \nl
ZW 1541+286      & IRAS 15418+2840    & 0.032  &        &   19.78  &   17.55    \nl

\tablenotetext{a}{2 pixels radius (or 0\farcs09)}
\tablenotetext{b}{11 pixels radius (or 0\farcs5), \nodata indicates central magnitudes are unavailabe due to saturation of central pixels}
\tablenotetext{}{ [....] denote upper limits on magnitude}

\enddata
\end{deluxetable}

\begin{deluxetable}{llcccclll}
\footnotesize
\tablenum{3}
\tablewidth{0pt}
\tablecaption{H II Galaxies}
\tablehead
{
\colhead{Galaxy} &
\colhead{Alternate} &
\colhead{12 $\mu$} &
\colhead{z} &
\colhead{Inner} &
\colhead{Outer} &
\colhead{Morphology} &
\colhead{} &
\colhead{Comments\tablenotemark {d}}
\\[.2ex]
\colhead{} &
\colhead{Designation} &
\colhead{} &
\colhead{} &
\colhead{Mag.\tablenotemark {a,c}} &
\colhead{Mag.\tablenotemark {b,c}} &
\colhead{MGT} &
\colhead{RC3} &
\colhead{}
}

\startdata

ESO 185-ig13   &                &     & 0.019 & 19.86 & 16.73 & ?      &        & CL, 	                    \nl
ESO 325-ig41   &                &     & 0.006 &\nodata&\nodata& Irr    &        & CL	                    \nl
ESO 350-ig38   &                &     & 0.020 &\nodata&\nodata& Irr    &        & CL, DC		            \nl
FRL 280          &                &     & 0.010 &\nodata&\nodata& SBb    & SBb    & B,CL                       \nl
G 1307-1608    &                &     & 0.032 & 20.80 & 18.79 & Irr    &        & faint nucleus, dwarf       \nl
G 1314-1532    &                &     & 0.013 & 23.96 & 21.15 & ?      &        & B                          \nl
IC 3576        &                &     & 0.003 &\nodata&\nodata& none   & Sm     & CL		            \nl
IC 4687        &                &     & 0.017 &\nodata&\nodata& Sd     & Sb     & interacting                \nl
KUV 13000+29   & KUV 13000+2908  &     & 0.020 & 21.72 & 18.92 &        &        &                            \nl
MRK 25        &                &     & 0.010 & 19.80 & 17.08 & Sb?    & E      & B                          \nl
MRK 52        &                &     & 0.007 &\nodata&\nodata& Sc     & SB0    &                            \nl
MRK 171       &                &     & 0.010 &\nodata&\nodata& Irr    &        & CL                         \nl
MRK 201       &                &     & 0.008 & 18.76 & 16.43 & Irr    & Im     & CL                         \nl
MRK 298       &                &     & 0.034 & 20.45 & 17.76 & no     & SA0    & DI	                    \nl
MRK 308       &                &     & 0.024 &\nodata&\nodata& S?     & Irr    & CL              	    \nl
MRK 567       &                &     & 0.033 & 20.14 & 17.25 & Sc     &        & flocculent                 \nl
MRK 703       &                &     & 0.013 &\nodata&\nodata& SBc    & Sab    & R,B                   	    \nl
MRK 759       &  NGC 4152         &     & 0.006 &\nodata&\nodata& SBd    & SABc   & DI,B                       \nl
MRK 789       &                &     & 0.032 & 19.96 & 17.15 & Irr    &        & collision                  \nl
MRK 930       &                &     & 0.019 &\nodata&\nodata& Irr    &        & CL                         \nl
MRK 1087      &                &     & 0.028 &\nodata&\nodata& ?      & S0     & D-E                        \nl
MRK 1133      &                &     & 0.024 & 19.23 & 17.01 & Sa     &        & bright central disk        \nl
MRK 1149      &                &     & 0.021 &\nodata&\nodata& SBc    &        & B,DC                       \nl
MRK 1261      &                &     & 0.026 & 20.30 & 17.38 & Sc/d   & Irr    & DC		            \nl
MRK 1308      &                &     & 0.004 &\nodata&\nodata& ?      & S0     & CL, DI                     \nl
MRK 1408      &                &     & 0.034 &\nodata&\nodata& SBb    &        & B	                    \nl
MRK 1414      &                &     & 0.014 &\nodata&\nodata& Sd     &        &                            \nl
MRK 1459      &                &     & 0.027 &\nodata&\nodata& Irr    &        & CL, collision              \nl
MRK 1490      &                &     & 0.026 &\nodata&\nodata& Sa     &        & R(partial)                 \nl
NGC 625$\star$   &                &     & 0.001 &\nodata&\nodata& Irr    & SBm    &                            \nl
NGC 1614         &                &     & 0.015 &\nodata&\nodata& Irr    & SBc    & CL,collision               \nl
NGC 2377         &                &     & 0.007 & 20.64 & 17.94 & Sd     & SAc    & R                          \nl
NGC 2989         &                &     & 0.013 &\nodata&\nodata& Sd     & SABbc  & flocculent, B?             \nl
NGC 3032         &                &     & 0.005 &\nodata&\nodata& Sa/b   & SAB0   & bright central disk        \nl
NGC 3310         &                &     & 0.003 &\nodata&\nodata& Sc/d   & SABbc  & CL, DC                     \nl
NGC 3353         &                &     & 0.003 &\nodata&\nodata& Irr    & Sb     & CL                         \nl
NGC 3504         &                &     & 0.005 &\nodata&\nodata& Sb     & Sab    & flocculent                 \nl
NGC 3738$\star$  &                &     & 0.001 &\nodata&\nodata& Irr    & Irr    & CL                         \nl
NGC 4694         &                &     & 0.004 &\nodata&\nodata& ?      & SB0    & CL                         \nl
NGC 4700$\star$  &                &     & 0.004 &\nodata&\nodata& Irr    & SBc    &                            \nl
NGC 4990         &  MRK 1344      &     & 0.010 &\nodata&\nodata& Sa     & S0     &                            \nl
NGC 5253$\star$  &                &     & 0.001 &\nodata&\nodata& Irr    & Irr    &                            \nl
NGC 5597         &                &     & 0.008 &\nodata&\nodata& Irr    & SABcd  & CL                         \nl
NGC 5757         &                &     & 0.008 & 19.85 & 17.03 & SBc    & SBb    & B,R, Sersic Pastoriza      \nl
NGC 7552         &                &     & 0.005 &\nodata&\nodata& Sd     & SBab   & B, flocculent              \nl
NGC 7714         &                &     & 0.009 &\nodata&\nodata& Ir     & SBb    & CL, collision              \nl
Q 1209-1105    &  [oh91] s08    &     & 0.016 &\nodata&\nodata& ?      &        & DI                         \nl
Q 1241+1624    &  [hb91] 1241+1 &     & 0.026 & 22.25 & 19.77 & ?      &        & dwarf                      \nl
SZ 80          &                &     & 0.026 & 20.71 & 18.80 & Irr    &        & bright nucleus             \nl
UGC 8929       &  WAS 87         &     & 0.027 &\nodata&\nodata& ?      &        & R                          \nl
WAS 96         &                &     & 0.034 & 21.56 & 18.49 & Irr    &        & CL                         \nl

\tablenotetext{a}{2 pixels radius (or 0\farcs09)}
\tablenotetext{b}{11 pixels radius (or 0\farcs5)}
\tablenotetext{c}{magnitudes only appear when there is no ambiguity as to the central point source}
\tablenotetext{d}
{
\nl
DI = irregular dust \nl
B = bar \nl
CL = cluster, lumpy HII region, knots \nl
DC = dust disk/dust lane passing close or through center (ie bisected nucleus) \nl
D-[direction] = dust lanes on one side of major axis, where direction is N, S, E, W, NW, NE, SW, or SE \nl
F/W = filaments/wisps \nl
R = ring \nl       
}
\tablenotetext{\star}{400x400 pixel images}

\enddata
\end{deluxetable}

\begin{deluxetable}{lcclll}
\footnotesize
\tablenum{4}
\tablewidth{0pt}
\tablecaption{Seyfert 1 Galaxies}
\tablehead
{
\colhead{Galaxy} &
\colhead{$\theta$} &
\colhead{$\epsilon$} &
\colhead{Morphology} &
\colhead{} &
\colhead{Comments\tablenotemark {\ast}}
\\[.2ex]
\colhead{} &
\colhead{} &
\colhead{} &
\colhead{MGT} &
\colhead{RC3} &
\colhead{}
}

\startdata

ESO 215-g14        &       &        &  Sa        &            &  SB0 (or tidal tail), faint galaxy\nl
ESO 323-g77        &       &        &  Sa/b      &            &  CL, asymmetric sprial arms      \nl
ESO 354-g4         &  31   & 0.14   &  Sa        &  Sb        &  normal                          \nl
ESO 362-g18        &       &        &  Sa        &  S0/a      &  spiral elongated nucleus, normal\nl
ESO 438-g9         &       &        &  SBc/d     &  Sab       &  B                               \nl
FRL 51               &  23   & 0.46   &  Sa        &  Sb        &  D-E                             \nl
FRL 1146             &  72   & 0.41   &  Sb        &            &  D-E                             \nl
H -1-0307-73       &       &        &  S(B)a     &            &  B                               \nl
H 1143-182         &       &        &  Irr       &            &  F/W, quasar-like                \nl
H 2106-099         &       &        &  L         &            &  galaxy not detectable, quasar-like\nl
IC 1816            &       &        &  SBa/b     &  Sab       &  R, B                            \nl
IC 4218            &       &        &  Sa        &  S?        &  DI                              \nl
IC 4329a           &       &        &  Sa        &  S0        &  DC, nearly edge-on              \nl
IRAS 1319-164        &       &        &  Sb?       &  SB?       &  edge of chip                    \nl
IRAS 13329-3402      &       &        &  S0        &            &  D-SW                            \nl
MCG 6-26-12        &       &        &  SB0       &  S?        &  B                               \nl
MCG 8-11-11        &       &        &  SB0       &  SB?       &  F/W, low surface brightness     \nl
MRK 6             &       &        &  S0        &  SB0       &  DI, E                           \nl
MRK 10            &  51   & 0.21   &  Sa/b      &  Sb        &  normal                          \nl
MRK 40            &       &        &  S0        &  S0        &  R, DI, smooth tidal tail        \nl
MRK 42            &       &        &  SBa       &  SB?       &  R, B, classic sersic/pastoriza  \nl
MRK 50            &       &        &  S0        &            &  DI                              \nl
MRK 79            &       &        &  SBc       &  SBb       &  F/W                             \nl
MRK 279           &       &        &  Sa        &  S0        &  normal Sa                       \nl
MRK 290           &       &        &  E         &  E         &  quasar-like                     \nl
MRK 334           &       &        &  Irr       &  Pec       &  CL, disturbed                   \nl
MRK 335           &       &        &  ?         &            &  F/W(1\arcsec to NE), quasar-like\nl
MRK 352           &  79   & 0.11   &  E         &  S0        &  smooth, low surface brightness  \nl
MRK 359           &       &        &  SBb/c     &  Pec       &  normal                          \nl
MRK 372           &       &        &  Sa        &            &  normal                          \nl
MRK 382           &       &        &  SBa       &  S?        &  B                               \nl
MRK 423           &       &        &  Sb        &  S0?       &  normal, some tidal disruption   \nl
MRK 471           &       &        &  SBc       &  SBb       &  B, DC, fragment to south        \nl
MRK 493           & 148   & 0.22   &  S(B)a     &  SBb       &  R                               \nl
MRK 516           &       &        &  Sc        &            &  D-N, double nucleus merger      \nl
MRK 530           &       &        &  Sa        &  Sb        &  D-NE, CL                        \nl
MRK 543           &       &        &  Sc        &  Im        &  normal                          \nl
MRK 590           &       &        &  Sa        &  Sa        &  normal                          \nl
MRK 595           &       &        &  Sa        &            &  D-SW, faint                     \nl
MRK 609           &       &        &  Sa/b      &  Im        &  flocculent?                     \nl
MRK 699           &       &        &  E         &            &  smooth, faint                   \nl
MRK 704           &       &        &  SBa       &  S?        &  B                               \nl
MRK 744           &       &        &  Sb        &  Sa        &  DI                              \nl
MRK 766           &       &        &  SBc       &  Sa        &  B, F/W, DI?                     \nl
MRK 817           &       &        &  SBc       &  S?        &  B                               \nl
MRK 833           &       &        &  Irr       &            &  CL, disturbed                   \nl
MRK 871           &       &        &  Sb        &  S?        &  D-N, DC                         \nl
MRK 885           &       &        &  SBb       &  S?        &  B                               \nl
MRK 896           &       &        &  Sc        &  S?        &  R(inclined)                     \nl
MRK 915           &       &        &  Sa        &  S?        &  F/W, DC, disturbed              \nl
MRK 1040          &       &        &  Sb        &  Sbc       &  DC, highly-inclined             \nl
MRK 1044          &       &        &  Sa        &  S?        &  Sersic/Pastoriza                \nl
MRK 1126          &       &        &  Sb        &  SBa       &  tight spiral arms nearly make a circle\nl
MRK 1218          &       &        &  SBa       &  S?        &  B                               \nl
MRK 1330          &       &        &  Sb/c      &  SBb       &  DC                              \nl
MRK 1376          &       &        &  edge on   &  Sa        &  DC, F/W                         \nl
MRK 1400          &  53   &  0.3   &  Sa        &            &  normal, highly inclined         \nl
MRK 1469          & 101   & 0.47   &  Sa        &            &  D-S, DC                         \nl
MS 1110+2210       &       &        &  E         &            &                                  \nl
NGC 235              & 119   & 0.36   &  Sa/b      &  S0        &  normal                          \nl
NGC 526a             &       &        &  E/S0      &            &  DI                              \nl
NGC1019             &  11   & 0.18   &  SBb       &  SBbc      &  R, B                            \nl
NGC 1566             &       &        &  Sb        &  SBbc      &  DC                              \nl
NGC 2639             &       &        &  Sb        &  Sa        &  DC, flocculent                  \nl
NGC 3227             &       &        &  ?         &  SBa       &  F/W, D-SW, dusty                \nl
NGC 3516             &       &        &  S0        &  S0        &  DI, B?                          \nl
NGC 3783             &       &        &  Irr       &  SBab      &  B                               \nl
NGC 4051             &       &        &  Sb        &  SBbc      &  D-SW                            \nl
NGC 4235             &       &        &  ?         &  Sa        &  DC                              \nl
NGC 4748             &       &        &  Sa        &            &  R                               \nl
NGC 5252             &  17   &  0.44  &  S0        &  S0        &  R(polar)                        \nl
NGC 5548             &       &        &  Sa        &  S0/a      &  normal                          \nl
NGC 5674             &       &        &  SBc       &  SBc       &  normal                          \nl
NGC 5940             &       &        &  SBc       &  SBab      &  B                               \nl
NGC 6104             &       &        &  SBb       &  S?        &  B, disturbed spiral pattern     \nl
NGC 6212             &       &        &  Sb        &  S?        &  flocculent spiral               \nl
NGC 6860             &       &        &  Sb        &  SBb       &  normal, D-S                     \nl
NGC 7213             &       &        &  Sa        &  Sa        &  normal                          \nl
NGC 7314             &       &        &  Sd        &  SBbc      &  DC                              \nl
NGC 7469             &       &        &  Sb/c      &  SBa       &  R                               \nl
PG 1310-108        &       &        &  ?         &            &  F/W, quasar-like                \nl
PKS 0518-458       &       &        &  E         &            &                                  \nl
TOL 1059+105       &       &        &  S0        &            &                                  \nl
TOL 2327-027       &       &        &  SB        &            &  CL, R                           \nl
UGC 1395           &       &        &  SBa       &  Sb        &  DI                              \nl
UGC 3223           &       &        &  S(B)b/c   &  SBa       &  D-N                             \nl
UGC 7064           &       &        &  Sa        &  S?        &  R                               \nl
UGC 10683b         &       &        &  SBa       &  S0        &  B                               \nl
UGC 12138          &       &        &  SBa       &  SBa       &  normal                          \nl
UM 614             & -11   & 0.29   &  S0        &            &  normal                          \nl
X 0459+034         &       &        &  E         &            &  F/W, smooth halo                \nl

\tablenotetext{\ast}
{
\nl
DI = irregular dust \nl
B = bar \nl
CL = cluster, lumpy HII region, knots \nl
DC = dust disk/dust lane passing close or through center (ie bisected nucleus) \nl
D-[direction] = dust lanes on one side of major axis, where direction is N, S, E, W, NW, NE, SW, or SE \nl
E/S0 = Elliptical \nl
F/W = filaments/wisps \nl
R = ring \nl       
}

\enddata
\end{deluxetable}

\begin{deluxetable}{lcclll}
\footnotesize
\tablenum{5}
\tablewidth{0pt}
\tablecaption{Seyfert 2 Galaxies}
\tablehead
{
\colhead{Galaxy} &
\colhead{$\theta$} &
\colhead{$\epsilon$} &
\colhead{Morphology} &
\colhead{} &
\colhead{Comments\tablenotemark {\ast}}
\\[.2ex]
\colhead{} &
\colhead{} &
\colhead{} &
\colhead{MGT} &
\colhead{RC3} &
\colhead{}
}

\startdata

ESO 137-g34        &       &         &  S0        &  SB0/a     &  CL, DI                               \nl
ESO 138-g01        &       &         &  L         &  E?        &  DI                                   \nl
ESO 139-g12        & 285   &  0.04   &  Sa        &  Sbc       &  normal                               \nl
ESO 353-g9         &       &         &  SBb       &  SBbc      &  DC, B                                \nl
ESO 362-g8         &       &         &  Sa        &  S0        &  DC                                   \nl
ESO 373-g29        &       &         &  Sb        &  SBab      &  DI, F/W, R, fragmentary dust tendril \nl
ESO 509-g66        &       &         &  ?         &            &  much of image is off chip            \nl
ESO 509-g66-c      &       &         &  S(B)a     &            &  normal spiral                        \nl
FRL 294              &       &         &  SBa       &  SB0       &  normal                               \nl
FRL 312              &       &         &  SBb       &  SBbc      &  B, inclined dust disk                \nl
FRL 315              &       &         &  SBa       &  SB0       &  R                                    \nl
FRL 316              &       &         &  S0        &  S0        &                                       \nl
FRL 334              &       &         &  S(B)b     &  SB0       &  B, DC                                \nl
FRL 341              &       &         &  Sa?       &  S0        &  D-NE                                 \nl
IC 184             &       &         &  SBb       &  SBa       &  B, inner + outer shells              \nl
IC 4870            &       &         &  ?         &  Im        &  F/W                                  \nl
IRAS 0147-076        &       &         &  E?        &            &  R(inner + outer)                     \nl
IRAS 0258-1136       &  50   &   0.3   &  S0/a      &  SBa       &  pretty smooth                        \nl
IRAS 0450+039        &       &         &  Sa        &            &  normal                               \nl
IRAS 0450-032        &       &         &  Sc        &            &  DC, two dust lanes not co-planar     \nl
IRAS 0457-756        & 217   &  0.25   &  SB0       &            &  B, old faded Sersic/Pastoriza        \nl
IRAS 1121-281        &       &         &  S0        &            &  smooth, very thin, part of           \nl
                  &       &         &            &            &  image is off chip                    \nl
IRAS 1305-241        &       &         &  Sc?       &            &  DC                                   \nl
IRAS 1443+272        &       &         &  SBa       &            &  B                                    \nl
IRAS 1548-037        &       &         &  S0        &            &  smooth                               \nl
IRAS 1832-594        &       &         &  Sa        &            &  F/W, point source                    \nl
IRAS 1833-654        &  28   & 0.38    &  E         &  S0?       &  DI                                   \nl
IRAS 2246-195        &       &         &  Sb        &            &  normal                               \nl
IRAS 2302-000        &       &         &  L         &            &  DI, R(broken)                        \nl
IRAS 2346+019        &       &         &  SBc       &            &  F/W, R, compact center               \nl
MCG -5-27-13       &       &         &  Sb        &  SBa       &  DC, F/W, dusty, bubbles along minor axis\nl
MRK 1             &  66   &  0.36   &  Sc        &  S         &  D-S, elongated east-west nucleus     \nl
MRK 3             &       &         &  S0        &  S0        &  F/W, bright linear center            \nl
MRK 176           &  51   &  0.45   &  Sa?       &  Sa        &  DI, nearly edge on, inner + outer rings\nl
MRK 198           &       &         &  Sa/b      &  SB0/a     &  R?                                   \nl
MRK 266sw         &       &         &  Irr       &  Pec       &  DI, collision                        \nl
MRK 270           &       &         &  E         &  S0        &  D-S, F/W                             \nl
MRK 313           &       &         &  Irr       &  SB0/a     &  DI, B, collision?                    \nl
MRK 348           &       &         &  S0        &  S0/a      &  double nuclei                        \nl
MRK 403           &       &         &  SB0/a     &            &  B                                    \nl
MRK 533           &  95   &  0.28   &  S(B)c     &  Sbc       &  F/W, R(partial)                      \nl
MRK 573           &       &         &  S0        &  SB0       &  F/W                                  \nl
MRK 577           &  54   &  0.25   &  S0        &  S0/a      &                                       \nl
MRK 607           &       &         &  Sb        &  Sa        &  D-SW                                 \nl
MRK 612           &       &         &  SB0/a     &  SB0/a     &  DC, flocculent, compact nucleus      \nl
MRK 620           &  67   &   0.5   &  SBb       &  SBa       &  F/W, center not extended along bar   \nl
MRK 622           &       &         &  S0        &  S?        &  DI                                   \nl
MRK 686           &       &         &  Sb        &  S?        &  D-SW, inclined spiral                \nl
MRK 917           &       &         &  Sc        &  SBa       &  CL                                   \nl
MRK 937           &       &         &  SBa/b     &            &  B, DC                                \nl
MRK 938           &       &         &  Sc        &  Pec       &  CL                                   \nl
MRK 955           &       &         &  S(B)b     &  S?        &  CL, Sersic/Pastoriza hot spot        \nl
MRK 993           &  29   & 0.31    &  Sb        &  Sa        &  D-SE, DC, highly-inclined            \nl
MRK 1058          & 108   &  0.41   &  Sb        &  S?        &  D-N, normal, inclined                \nl
MRK 1066          &       &         &  Sc        &  SB0       &  F/W, dusty                           \nl
MRK 1073          &       &         &  Sc        &  SBb       &  F/W, Irregular                       \nl
MRK 1157          &       &         &  Sb/c      &  SB0/a     &  R                                    \nl
MRK 1193          &       &         &  SBb       &            &  B, F/W                               \nl
MRK 1210          &       &         &  Sa        &  S?        &  normal                               \nl
MRK 1370          &  34   &   0.3   &  Sa        &            &  B(north-south)                       \nl
NGC 424              &       &         &  Sb        &  S0/a      &  D-S                                  \nl
NGC 513              &       &         &  Sb/c      &  S?        &  normal                               \nl
NGC 788              & 112   & .1-.3   &  S0        &  S0/a      &  DI, F/W                              \nl
NGC 1125             &       &         &  Sb/c      &  SB0/a     &  DC                                   \nl
NGC 1144             &       &         &  Irr       &  Ring B    &  DI, collision                        \nl
NGC 1241             &       &         &  Sb/c      &  SBb       &  normal                               \nl
NGC 1358             &       &         &  SB0       &  SB0/a     &  B, DI                                \nl
NGC 1386             &       &         &  Sb/c      &  SB0       &  DC                                   \nl
NGC 1410             &       &         &  S0        &  E pec     &  F/W, collision: polar dust ring      \nl
NGC 1667             &       &         &  Sc        &  SBc       &  DI                                   \nl
NGC 2110             &       &         &  Sa        &  SB0       &  D-W                                  \nl
NGC 2992             &       &         &  ?         &  Sa        &  DC                                   \nl
NGC 3081             &       &         &  SB0/a     &  SB0/a     &  DI                                   \nl
NGC 3362             &       &         &  Sb        &  SBc       &  F/W, hook-like protrusion from nucleus\nl
NGC 3393             &       &         &  Sa        &  SBa       &  F/W, like MRK 573                    \nl
NGC 3982             &       &         &  Sb/c      &  SBb       &  normal                               \nl
NGC 4156             &  63   &  0.27   &  SBa/b     &  SBb       &  DI                                   \nl
NGC 4507             &       &         &  S(B)a/b   &  SBb       &  D-SE, B, twisted isophote            \nl
NGC 4922b            & 148   &  0.12   &  E         &            &  DI, dust splotches                   \nl
NGC 4939             &       &         &  Sa        &  Sbc       &  F/W, D-W                             \nl
NGC 4968             &       &         &  Sa        &  SB0/a     &  DI                                   \nl
NGC 5135             &       &         &  Sc        &  SBab      &  CL                                   \nl
NGC 5347             &       &         &  Sb        &  SBab      &  normal                               \nl
NGC 5427             &       &         &  Sc        &  Sc        &  normal                               \nl
NGC 5929             &       &         &  S0        &  Sab       &  DC                                   \nl
NGC 5953             &       &         &  Sc        &  Sa        &  flocculent                           \nl
NGC 5995             &       &         &  S(B)c     &            &  normal                               \nl
NGC 6211             &  40   &  0.32   &  E         &  SB0/a     &  shell                                \nl
NGC 6217             &       &         &  Sc        &  SBbc      &  CL, DC                               \nl
NGC 6221             &       &         &  Sd        &  SBc       &  CL, Irregular                        \nl
NGC 6300             &       &         &  Sd        &  SBb       &  DI                                   \nl
NGC 6393             &       &         &  S(B)c     &  SBb       &  DC                                   \nl
NGC 7130             &       &         &  Sd        &  Sa        &  CL, DC                               \nl
NGC 7172             &       &         &  ?         &  Sa        &  DC                                   \nl
NGC 7212             &       &         &  Irr?      &  S?        &  F/W, DI                              \nl
NGC 7319             &       &         &  Irr?      &  SBbc      &  F/W, DI, shells                      \nl
NGC 7410             &       &         &  E/S0      &  SBa       &  DC                                   \nl
NGC 7582             &       &         &  ?         &  SBab      &  DC, F/W, inclined dust disk          \nl
NGC 7590             &       &         &  Sd        &  Sbc       &  DC                                   \nl
NGC 7592             &       &         &  Irr       &  S0        &  CL, Irregular                        \nl
NGC 7682             &       &         &  SB0       &  SBab      &  CL, F/W, early type                  \nl
NGC 7743             &       &         &  S0        &  SBab      &  F/W, DI, early type                  \nl
PKS 2048-572       &  81   & 0.68    &  E?        &  S0        &  DC, F/W(turbulent)                   \nl
PKS 2158-380       &       &         &  Sa        &  SB0       &  D-N                                  \nl
Q 1234+0848        &       &         &  Irr       &  SBb       &  dwarf                                \nl
UGC 3255           &       &         &  Sb/c      &  S?        &  DC                                   \nl
UGC 4332           &       &         &            &  Sa        &  DC                                   \nl
UGC 6100           &       &         &  Sb        &  Sa?       &  normal                               \nl
UM 105             &       &         &  Sc?       &            &  DI                                   \nl
UM 319             &       &         &  Sc        &            &  CL, dusty                            \nl
UM 625             &       &         &  S0        &            &  R(partial)                           \nl
WAS 2              &       &         &  SB0       &            &  B                                    \nl
ZW 1408+137        &       &         &  ?         &            &  DI, disturbed dust                   \nl
ZW 1541+286        &       &         &  SBa       &            &  B, lens                              \nl

\tablenotetext{\ast}
{
\nl
DI = irregular dust \nl
B = bar \nl
CL = cluster, lumpy HII region, knots \nl
DC = dust disk/dust lane passing close or through center (ie bisected nucleus) \nl
D-[direction] = dust lanes on one side of major axis, where direction is N, S, E, W, NW, NE, SW, or SE \nl
E/S0 = Elliptical \nl
F/W = filaments/wisps \nl
R = ring \nl       
}

\enddata
\end{deluxetable}

\begin{deluxetable}{lllccccc}
\footnotesize
\tablenum{1}
\tablewidth{0pt}
\tablecaption{Seyfert 1 Galaxies}
\tablehead
{
\colhead{Galaxy} &
\colhead{Alternate} &
\colhead{z} &
\colhead{12\micron} &
\colhead{Inner} &
\colhead{Outer} &
\colhead{Seyfert} &
\colhead{Saturated}
\\[.2ex]
\colhead{} &
\colhead{Designation} &
\colhead{} &
\colhead{} &
\colhead{Mag.\tablenotemark{a}} &
\colhead{Mag.\tablenotemark{b}} &
\colhead{Class} &
\colhead{Unsaturated, Resolved}
}

\startdata

ESO 215-g14    &                 & 0.019   &     & 17.56   & 15.92   &   1   & US1 \nl
ESO 323-g77    &                 & 0.015   &     & 16.69   & \nodata &   1   & SS1 \nl
ESO 354-g4     &                 & 0.033   &     & 18.42   & 16.66   &   1   & US1 \nl
ESO 362-g18    & MCG -5-13-17     & 0.013   & X   & \nodata & \nodata &   1   & RS1 \nl
ESO 438-g9     & IRAS 11083-2813  & 0.024   &     & 17.44   & \nodata & 1.5   & SS1 \nl
FRL   51       & ESO 140-g43      & 0.014   &     & 17.17   & \nodata &   1   & SS1 \nl
FRL 1146       &                 & 0.032   &     & 16.54   & 14.40   & 1.5   & RS1 \nl
H -1-0307-73   & H0307-730       & 0.028   &     & 18.29   & \nodata &       & SS1 \nl
H 1143-182     &                 & 0.033   &     & 17.16   & \nodata &   1   & SS1 \nl
H 2106-099     &                 & 0.027   &     & 16.87   & \nodata & 1.2   & SS1 \nl
IC 1816        & ESO 355-g25      & 0.017   &     & 18.10   & \nodata &   1   & RS1 \nl
IC 4218        & UGC 8348         & 0.019   &     & 18.68   & 17.41   &   1   & US1 \nl
IC 4329a       & ESO 445-g50      & 0.016   & X   & 16.91   & \nodata &   1   & SS1 \nl
IRAS 1319-164  & MCG -3-34-63     & 0.017   & X   & 16.43   & \nodata & 1.8   & RS1 \nl
IRAS 13329-340 & MCG -6-30-15     & 0.008   & X   & 17.26   & \nodata &   1   & SS1 \nl
MCG 6-26-12    & KUG 1136+342     & 0.032   &     & 18.43   & 17.52   & 1.5   & US1 \nl
MCG 8-11-11    & UGC 3374         & 0.020   &     & 17.07   & \nodata & 1.5   & SS1 \nl
MRK    6          & IC 0450          & 0.019   & X   & 16.77   & \nodata & 1.5   & SS1 \nl
MRK   10         & UGC 4013         & 0.030   &     & 18.46   & 16.85   &   1   & US1 \nl
MRK   40         & ARP 151          & 0.020   &     & 18.68   & 17.03   &   1   & US1 \nl
MRK   42         &                 & 0.024   &     & 19.01   & 17.54   &   1   & RS1 \nl
MRK   50         &                 & 0.023   &     & 18.35   & 17.12   &   1   & US1 \nl
MRK   79         & UGC 3973         & 0.022   & X   & 16.92   & \nodata & 1.2   & SS1 \nl
MRK  279        & UGC 8823         & 0.031   &     & 16.66   & \nodata &   1   & SS1 \nl
MRK  290        &                 & 0.029   &     & 17.65   & \nodata &   1   & SS1 \nl
MRK  334        & UGC 6            & 0.022   &     & 17.99   & \nodata & 1.8   & RS1 \nl
MRK  335        & PG 0003+199      & 0.025   & X   & 16.51   & \nodata &   1   & SS1 \nl
MRK  352        &                 & 0.015   &     & 18.53   & 17.04   &   1   & US1 \nl
MRK  359        & UGC 1032         & 0.017   &     & 17.97   & \nodata & 1.5   & SS1 \nl
MRK  372        & IC 1854          & 0.031   &     & 18.70   & 16.88   & 1.5   & RS1 \nl
MRK  382        &                 & 0.034   &     & 18.44   & \nodata &   1   & SS1 \nl
MRK  423        & MCG 6-25-72      & 0.032   &     & 18.96   & 16.49   & 1.9   & RS1 \nl
MRK  471        & UGC 9214         & 0.034   &     & 19.42   & 17.78   & 1.8   & RS1 \nl
MRK  493        & UGC 10120        & 0.031   &     & 18.67   & 17.13   &   1   & US1 \nl
MRK  516        & IRAS 21538+0707  & 0.028   &     & 17.70   & 16.48   & 1.8   & RS1 \nl
MRK  530        & NGC 7603           & 0.029   & X   & 16.97   & \nodata & 1.5   & SS1 \nl
MRK  543        & NGC 7811           & 0.026   &     & 18.42   & \nodata &   1   & SS1 \nl
MRK  590        & NGC 863            & 0.027   &     & 16.76   & \nodata & 1.2   & SS1 \nl
MRK  595        &                 & 0.028   &     & 18.56   & 16.48   &   1   & RS1 \nl
MRK  609        & IRAS 03229-0618  & 0.032   &     & 17.48   & \nodata & 1.8   & RS1 \nl
MRK  699        & IIIZW77         & 0.034   &     & 18.42   & 17.11   & 1.2   & US1 \nl
MRK  704        &                 & 0.029   & X   & 17.45   & \nodata & 1.5   & SS1 \nl
MRK  744        & NGC 3786           & 0.010   &     & 17.05   & \nodata & 1.8   & SS1 \nl
MRK  766        & NGC 4253           & 0.012   & X   & 17.13   & \nodata & 1.5   & SS1 \nl
MRK  817        & UGC 9412         & 0.033   & X   & 17.43   & \nodata & 1.5   & SS1 \nl
MRK  833        &                 & 0.039   &     & 22.46   & 19.33   &       & RS1 \nl
MRK  871        & IC 1198          & 0.034   &     & 18.61   & \nodata & 1.5   & SS1 \nl
MRK  885        &                 & 0.026   &     & 19.95   & 17.80   & 1.5   & RS1 \nl
MRK  896        &                 & 0.027   &     & 18.30   & \nodata &   1   & US1 \nl
MRK  915        & MCG -2-57-23     & 0.025   &     & 19.11   & 17.10   &   1   & RS1 \nl
MRK 1040       & NGC 931            & 0.016   & X   & 18.46   & \nodata & 1.5   & SS1 \nl
MRK 1044       &                 & 0.016   &     & 17.15   & \nodata &   1   & SS1 \nl
MRK 1126       & NGC 7450           & 0.010   &     & 18.75   & 16.80   & 1.5   & RS1 \nl
MRK 1218       & NGC 2622           & 0.028   &     & 18.66   & 16.82   & 1.8   & RS1 \nl
MRK 1330       & NGC 4593           & 0.009   & X   & 17.11   & \nodata &   1   & SS1 \nl
MRK 1376       & NGC 5506           & 0.007   & X   & 19.41   & 16.99   & 1.9   & RS1 \nl
MRK 1400       & CGCG 414-001     & 0.029   &     & 20.07   & 17.54   &   1   & RS1 \nl
MRK 1469       & MCG 9-20-136     & 0.031   &     & 18.98   & 17.18   & 1.5   & RS1 \nl
MS 1110+2210   & MS 1110.3+2210   & 0.030   &     & 20.74   & 18.37   &   1   & RS1 \nl
NGC  235        &                 & 0.022   &     & 18.72   & 16.16   &   1   & RS1 \nl
NGC  526a       &                 & 0.018   & X   & 17.98   & \nodata & 1.5   & RS1 \nl
NGC 1019       &                 & 0.024   &     & 18.75   & 17.11   &   1   & US1 \nl
NGC 1566       &                 & 0.004   & X   & 16.84   & \nodata &   1   & RS1 \nl
NGC 2639       &                 & 0.011   & X   & 19.41   & 16.33   &   1   & RS1 \nl
NGC 3227       & UGC 5620         & 0.003   & X   & 16.33   & \nodata & 1.5   & SS1 \nl
NGC 3516       & UGC 6153         & 0.009   & X   & 15.83   & \nodata & 1.5   & SS1 \nl
NGC 3783       & ESO 378-g14      & 0.009   &     & 17.43   & \nodata &   1   & SS1 \nl
NGC 4051       &                 & 0.002   & X   & 16.28   & \nodata &   1   & SS1 \nl
NGC 4235       & UGC 7310         & 0.007   &     & 18.60   & 16.78   &   1   & RS1 \nl
NGC 4748       & IRAS 12495-1308  & 0.014   & X   & 17.57   & \nodata &   1   & SS1 \nl
NGC 5252       &                 & 0.022   &     & 19.47   & 17.08   & 1.9   & RS1 \nl
NGC 5548       & MRK 1509         & 0.017   & X   & 16.55   & \nodata & 1.5   & SS1 \nl
NGC 5674       &                 & 0.025   &     & 19.12   & 16.77   & 1.9   & RS1 \nl
NGC 5940       & UGC 9876         & 0.033   &     & 18.75   & 17.30   &   1   & US1 \nl
NGC 6104       &                 & 0.028   &     & 19.88   & 18.01   & 1.5   & RS1 \nl
NGC 6212       &                 & 0.030   &     & 20.03   & 17.48   &   1   & RS1 \nl
NGC 6860       &                 & 0.015   & X   & 17.37   & \nodata &   1   & SS1 \nl
NGC 7213       &                 & 0.006   & X   & 16.85   & \nodata &   1   & US1 \nl
NGC 7314       & ESO 533-g53      & 0.006   & X   & 19.75   & 18.23   & 1.9   & RS1 \nl
NGC 7469       & MRK 1514         & 0.017   & X   & 16.38   & \nodata &   1   & SS1 \nl
PG 1310-108    & IISZ10          & 0.034   &     & 18.00   & \nodata &       & SS1 \nl
PKS 0518-458   &                 & 0.034   &     & 18.82   & 17.41   &       & US1 \nl
TOL 1059+105   &                 & 0.034   &     & 19.26   & 17.64   &   1   & US1 \nl
TOL 2327-027   & UM 163           & 0.033   &     & 19.59   & 17.75   &   1   & US1 \nl
UGC  1395       & UM 146           & 0.017   &     & 19.69   & 17.46   & 1.9   & US1 \nl
UGC  3223       &                 & 0.018   &     & 18.76   & 17.14   &   1   & RS1 \nl
UGC  7064       & WAS 45           & 0.024   & X   & 18.40   & 15.98   & 1.9   & RS1 \nl
UGC 10683b     &                 & 0.031   &     & 18.49   & 17.18   &   1   & US1 \nl
UGC 12138      & IRAS 22377+0747  & 0.025   &     & 17.67   & \nodata & 1.8   & US1 \nl
UM 614         &                 & 0.033   &     & 18.60   & 16.69   &   1   & US1 \nl
X 0459+034     & MS 0459.5+0327   & 0.016   &     & 16.43   & \nodata &   1   & RS1 \nl

\tablenotetext{a}{2 pixels radius (or 0\farcs09)}
\tablenotetext{b}{11 pixels radius (or 0\farcs5), 
\nodata indicates no central magnitude available due to saturation of central
pixels} 

\enddata
\end{deluxetable}